\documentclass[a4paper,11pt]{article}
\usepackage{jcappub} % for details on the use of the package, please see the JINST-author-manual
\usepackage{lineno}
\usepackage{tablefootnote}
\usepackage{threeparttable}
\usepackage{graphicx}% Include figure files
\usepackage{dcolumn}% Align table columns on decimal point
\usepackage{bm}% bold math
\usepackage{gensymb}
\usepackage{multirow}
\usepackage{booktabs}

\usepackage{tablefootnote}
\usepackage{hyperref}
\usepackage[utf8]{inputenc}
\usepackage[T1]{fontenc}
\usepackage{mathptmx}
\usepackage{etoolbox}
\usepackage{xcolor}
\usepackage{cprotect}
\usepackage{aas_macros}
\usepackage{mathtools}

\DeclarePairedDelimiter\floor{\lfloor}{\rfloor}

\usepackage{scalerel}
\usepackage{tikz}
\usetikzlibrary{svg.path}
\definecolor{orcidlogocol}{HTML}{A6CE39}
\tikzset{
  orcidlogo/.pic={
    \fill[orcidlogocol] svg{M256,128c0,70.7-57.3,128-128,128C57.3,256,0,198.7,0,128C0,57.3,57.3,0,128,0C198.7,0,256,57.3,256,128z};
    \fill[white] svg{M86.3,186.2H70.9V79.1h15.4v48.4V186.2z}
                 svg{M108.9,79.1h41.6c39.6,0,57,28.3,57,53.6c0,27.5-21.5,53.6-56.8,53.6h-41.8V79.1z M124.3,172.4h24.5c34.9,0,42.9-26.5,42.9-39.7c0-21.5-13.7-39.7-43.7-39.7h-23.7V172.4z}
                 svg{M88.7,56.8c0,5.5-4.5,10.1-10.1,10.1c-5.6,0-10.1-4.6-10.1-10.1c0-5.6,4.5-10.1,10.1-10.1C84.2,46.7,88.7,51.3,88.7,56.8z};
  }
}
\newcommand\orcidlink[1]{\href{https://orcid.org/#1}{\mbox{\scalerel*{
\begin{tikzpicture}[yscale=-1,transform shape]
\pic{orcidlogo};
\end{tikzpicture}
}{|}}}}

\title{SPECTER: An Instrument Concept for CMB Spectral Distortion Measurements with Enhanced Sensitivity}
\author{Alina Sabyr$^{a,b,c}$\,\orcidlink{0000-0003-3595-4384}}
\affiliation{$^{a}$Department of Astronomy, Columbia University, New York, NY 10027, USA}
\affiliation{$^{b}$Berkeley Center for Cosmological Physics, Department of Physics, University of California, Berkeley, CA 94720, USA}
\affiliation{$^{c}$Lawrence Berkeley National Laboratory, One Cyclotron Road, Berkeley, CA 94720, USA}
\emailAdd{as6131@columbia.edu}
\author{Carlos Sierra$^{d,e,f}$\,\orcidlink{0000-0002-9246-5571}}
\affiliation{$^{d}$Kavli Institute for Particle Astrophysics and Cosmology, Stanford University, Stanford, CA 94305, USA}
\affiliation{$^{e}$Department of Physics, University of Chicago, Chicago, IL 60637, USA}
\affiliation{$^{f}$Kavli Institute for Cosmological Physics, University of Chicago, Chicago, IL 60637, USA}
\emailAdd{csierra@stanford.edu}
\author{J.~Colin Hill$^{g}$\,\orcidlink{0000-0002-9539-0835}} 
\affiliation{$^{g}$Department of Physics, Columbia University, New York, NY 10027, USA}
\emailAdd{jch2200@columbia.edu}
\author{Jeffrey J.~McMahon$^{h, e, f}$}
\affiliation{$^{h}$Department of Astronomy and Astrophysics, University of Chicago, Chicago, IL 60637, USA}
\emailAdd{jjm@uchicago.edu}

\abstract{Deviations of the cosmic microwave background (CMB) energy spectrum from a perfect blackbody uniquely probe a wide range of physics, ranging from fundamental physics in the primordial Universe ($\mu$-distortion) to late-time baryonic feedback processes ($y$-distortion). While the $y$-distortion can be detected with a moderate increase in sensitivity over that of \emph{COBE/FIRAS}, the $\Lambda$CDM-predicted $\mu$-distortion is roughly two orders of magnitude smaller and requires substantial improvements, with foregrounds presenting a serious obstacle.  Within the standard model, the dominant contribution to $\mu$ arises from energy injected via Silk damping, yielding sensitivity to the primordial power spectrum at wavenumbers $k \approx 1-10^{4}$ Mpc$^{-1}$. Here, we present a new instrument concept, \emph{SPECTER}, with the goal of robustly detecting $\mu$. The instrument technology is similar to that of \emph{LiteBIRD}, but with an absolute temperature calibration system. Using a Fisher approach, we optimize the instrument's configuration to target $\mu$ while marginalizing over foreground contaminants. Unlike Fourier-transform-spectrometer-based designs, the specific bands and their individual sensitivities can be independently set in this instrument, allowing significant flexibility. We forecast \emph{SPECTER} to observe the $\Lambda$CDM-predicted $\mu$-distortion at $\approx 5\sigma$ (10$\sigma$) assuming an observation time of 1 (4) year(s) (corresponding to mission duration of 2 (8) years), after foreground marginalization. Our optimized configuration includes 16 bands spanning 1-2000 GHz with $\sim$degree-scale angular resolution at $\sim150$ GHz and 1100 total detectors. \emph{SPECTER} will additionally measure the $y$-distortion at sub-percent precision and its relativistic correction at percent-level precision, yielding tight constraints on the total thermal energy and mean temperature of ionized gas.}
\begin{document}
\maketitle
\section{Introduction}\label{sec:intro}
The cosmic microwave background (CMB) has the most perfect blackbody spectrum measured in nature, as was observed by \emph{COBE/FIRAS} in the 1990s \cite{Mather1990, Mather1994, Wright1994, Fixsen1996}. Nevertheless, small deviations --- referred to as CMB spectral distortions --- can arise both from exotic and standard physical processes, which drive matter and radiation out of thermal equilibrium at redshifts up to two million. Measurements of these signals would constrain early-Universe physics and provide unique opportunities to search for evidence of non-standard models including the decay of new particles \cite{HuSilk1993, McDonald2000}, black hole evaporation \cite{Carr2010, Nakama2018}, and many other elements of new physics (via the $\mu$-distortion). In particular, constraints on the sky-averaged monopole $\mu$-distortion probe inflationary physics, independent from and complementary to B-mode polarization searches, since the known and dominant source of this signal predicted within $\Lambda$CDM is the dissipation of small-scale primordial acoustic modes \cite{Sunyaev1970,Daly1991,Hu1994a,ChlubaKhatriSunyaev2012,KhatriSunyaevChluba2012, ChlubaErickcekBenDayan2012}. In addition to offering a unique window into the early Universe, spectral distortions also provide a measure of the mean ionized gas properties in the Universe via the integral of the thermal Sunyaev-Zel'dovich effect (the $y$-distortion). This would give us a new insight into the astrophysics of the late-time Universe, delivering powerful constraints relevant to galaxy formation and evolution models and thus the largely unknown physics of baryonic feedback \cite{Zeldovich1969,Sunyaev1970, Hill2015, Thiele2022} (see, e.g., Refs.~\cite{decadal2019, Voyage2050} for a more detailed description of the science case for CMB spectral distortions).

Measuring the $y$- and $\mu$-distortions requires an absolutely calibrated measurement of the CMB energy spectrum, with instrumental noise and systematic effects controlled respectively to a factor of 10 and 10,000 times better than the current state-of-the art measurement from \emph{COBE/FIRAS}, which was developed in the 1980s. Since then, the detector technologies used for measurements of the CMB intensity have advanced enormously, effectively following an analog of Moore's law, which makes the required improvements over \emph{COBE} possible. Several teams have proposed developing a new version of the \emph{COBE/FIRAS} instrument based on an improved Fourier Transform Spectrometer (FTS) coupled to modern detectors (e.g., \emph{PIXIE} \cite{Pixie,pixie2024}, \emph{Super-PIXIE} \cite{super-pixie}, \emph{PRISM} \cite{Prism}, and \emph{Voyage 2050} \cite{Voyage2050}). Since the $y$-distortion signal is predicted to be only about an order of magnitude below the \emph{COBE/FIRAS} noise levels~\cite{Hill2015,Dolag2016}, a single satellite of this design could easily measure the $y$-distortion. On the other hand, the $\Lambda$CDM $\mu$-distortion signal is roughly 100 times smaller than the $y$-distortion, requiring higher sensitivity and much more stringent foreground mitigation to enable a detection~\cite{Chluba2013a}.  In particular, the complexity of low-frequency ($\lesssim 100$ GHz) foregrounds renders $\mu$ particularly challenging to detect~\cite{Abitbol2017}, and likely necessitates the use of multiple FTS instruments covering different frequency ranges in order to have sufficient post-foreground-marginalization sensitivity~\cite{super-pixie,Voyage2050}.  In addition to satellite mission proposals, there are also ground-based (e.g., \emph{COSMO} \cite{Cosmo}, \emph{TMS} \cite{TMS}) and balloon-borne (e.g., \emph{BISOU} \cite{Bisou}) experiments that are currently being developed, most targeting mainly a $y$-distortion measurement due to the signal's much larger amplitude.

One of the main observational and data analysis challenges is the presence of astrophysical foregrounds, both of Galactic and extragalactic origin, across all frequencies of interest. Most of the foreground emission spectra are not  constrained to the sufficient level of precision for a $\mu$-distortion detection ($\approx$ 1-10 nK), and, therefore, future spectral distortion missions must have sufficient sensitivity to place constraints on the distortion signals while simultaneously marginalizing over each of the foregrounds. For example, Ref.~\cite{Abitbol2017} (hereafter A17) showed that an extended \emph{PIXIE} mission would not be able to detect the $\mu$-distortion when marginalizing over all foreground components, despite having instrumental noise that is formally lower than the targeted $\mu$ signal.

In single-FTS-based space mission set-ups, the spectral resolution and noise levels are nearly uniform across all frequency channels~\cite{Pixie}. In order to substantially improve the overall sensitivity, the mission duration, detector count, or \'{e}tendue (the product of detector area and solid angle) must be increased, any of which would raise the mission's cost significantly. The total foreground contribution, however, varies significantly across frequencies, with the highest contamination at high and low frequencies  where dust and synchrotron emission dominate, respectively. This situation motivates alternative instrument concepts, with the flexibility needed to optimize the instrumental noise as a function of frequency in order to optimally mitigate foregrounds while maximizing spectral distortion sensitivity.

In this work we present an instrument concept for a satellite mission, the Spectral Photometry Experiment for Cosmic ThErmal distoRtions (\emph{SPECTER}). This instrument is similar to the \emph{LiteBIRD} \cite{LiteBird2023} and \emph{PICO} \cite{PICO2019} mission concepts, but with the introduction of a robust absolute calibration system.\footnote{Previously, Refs.~\cite{Mukherjee2018, Mukherjee2019} studied the possibility of measuring spectral distortions with an imager mission together with an absolute calibrator and an interfrequency calibrator.} The advantage of this approach, compared to an FTS, is that the precise bands and the sensitivity within each band can be determined independently. We can tailor the exact instrument configuration to a science target, which in this case is the $\mu$-distortion, and the sensitivity of each band in this type of design is set only by its own photon loading. In this paper, we use a Fisher forecast to find a (near-)optimal instrument configuration for \emph{SPECTER} to observe the $\mu$-distortion, while marginalizing over astrophysical foregrounds and keeping the number of detectors of the instrument relatively low. 

The remainder of this paper is organized as follows.  In Sec.~\ref{sec:CMBSD} we review CMB spectral distortions and the models that we use in the forecasts. We describe the detector sensitivity estimates in Sec.~\ref{sec:bolocalc}, instrument configuration optimization in Sec.~\ref{sec:configuration_optimization}, absolute calibration requirements in Sec.~\ref{sec:calibration}, and the impact of the fiducial sky model assumptions in Sec.~\ref{sec:sky_modeling_choices}. We discuss the final instrument design in Sec.~\ref{sec:SPECTER} and conclude in Sec.~\ref{sec:conclusion}.
 
\section{CMB Spectral Distortions}
\label{sec:CMBSD}
CMB spectral distortions can be generated by various physical processes in both the early- and late-time Universe. At redshifts $z \gtrsim 2 \times 10^{6}$, Compton scattering, double-Compton, and bremsstrahlung emission maintain thermal equilibrium between matter and radiation \cite{Zeldovich1969,SunyaevZeldovich1970, Illarionov1975a, Illarionov1975b, Weymann1965, Lightman1981, Danese1982}. Any possible distortions at those redshifts are erased and a blackbody spectrum is preserved, simply with an overall (unobservable) temperature shift.

Due to the expansion and the resulting cooling of the Universe, the double-Compton and bremsstrahlung rates become lower than the Hubble rate at $z \lesssim 2 \times 10^6$, i.e., these processes become inefficient.  Only Compton scattering remains efficient at $5 \times 10^{4} \lesssim z \lesssim 2 \times 10^{6}$, sustaining a kinetic equilibrium. Energy release during this period leads to a Bose-Einstein spectrum characterized by a non-zero chemical potential $\mu$ or a $\mu$-type distortion. Establishing this equilibrium distribution requires that Compton scattering is efficient, so a Bose-Einstein spectrum cannot be formed at lower redshifts. The primordial $\mu$- (or chemical potential) distortion thus offers a direct and clean probe of the Universe during the pre-recombination era\footnote{Recently, Ref.~\cite{Chluba2024dark_photons} has shown that photon to dark photon conversions can produce a spectral distortion very similar to $\mu$-type after recombination, but this is a beyond-standard-model process.}). The standard sources of this distortion are energy injected by the Silk damping of small-scale primordial acoustic waves \cite{Sunyaev1970,Daly1991,BarrowColes1991} and adiabatic cooling of the CMB due to interactions with electrons \cite{Chluba2005, ChlubaSunyaev2012} (the latter being sub-dominant).\footnote{Note that the electron and photon temperatures scale respectively as $(1+z)^2$ and $(1+z)$ in the absence of interactions, and thus the electrons are continually cooling the photons during this era.}  Thus, the $\mu$-distortion is sensitive to the primordial power spectrum at $k\sim 1-10^{4}$ Mpc$^{-1}$ and can be used as a robust, independent probe of inflationary scenarios (\cite[e.g.,][]{ChlubaErickcekBenDayan2012,Dent2012,KhatriSunyaev2013,Clesse2014, Cabass2016}) on scales inaccessible to other observations. 

At $z \lesssim 5 \times 10^{4}$, Compton scattering also ceases to be efficient. During this epoch, a $y$-type distortion is formed if energy or entropy is injected into the plasma. The known source of this distortion is the inverse-Compton scattering of CMB photons off of energetic, free electrons in the intergalactic (IGM) and intracluster media (ICM) during the epoch of reionization and structure formation --- also known as the thermal Sunyaev-Zel'dovich (tSZ) effect~\cite{Zeldovich1969,Sunyaev1970}. The strength of this distortion is parameterized by the Compton-$y$ parameter, which is proportional to the electron pressure along the line of sight and therefore gives an integral constraint on the thermal energy stored in electrons responsible for the scattering. Measurement of the relativistic correction to the $y$-distortion (which is non-negligible due to the high temperatures that the ICM can reach, $T \sim 10^7$ K), gives a constraint on the mean electron temperature. Thus, placing tight constraints on the $y$-distortion uniquely characterizes the baryons in the observable Universe and has the potential to give powerful astrophysical constraints on galaxy formation \cite{Hill2015,Thiele2022}. While a $y$-type distortion can also be generated by the dissipation of acoustic waves prior to recombination (but at $z \lesssim 5 \times 10^4$), the distortion signal generated post-recombination~\cite{Hill2015} is roughly three orders of magnitude larger than this Silk-damping signal~\cite{ChlubaKhatriSunyaev2012}. Hence, the $y$-distortion primarily probes the late-time Universe. 

Together, the $\mu$- and $y$-distortions trace the thermal history of the Universe, from long before recombination through the epoch of structure formation. In this work, we use distortion models from A17 (described below), where only sources within the standard cosmological model are considered. However, it is important to note that $\mu$- and $y$-type distortions can also be produced by non-standard sources, such as primordial black holes \cite{Carr2010,AliHaimoudKamionkowski2017,Nakama2018}, dark matter annihilation and decay \cite{McDonald2000}, cosmic strings \cite{Ostriker1987, Tashiro2013,Cyr2023cosmic_strings}, and primordial magnetic fields \cite{Jedamzik2000,Kunze2014,Wagstaff2015}.  Thus, high-precision measurements of these signals represent an exciting discovery space for new physics.

Other types of spectral distortions are also predicted to exist in our Universe, including the cosmological recombination radiation (CRR) \cite{Zeldovich1968crr,Dubrovich1975,SunyaevChluba2009CRR,ChlubaAliHaimoud2016} and a ``residual'' or $r$-type distortion from the gradual transition between $\mu$- and $y$-type eras around $z \sim 5 \times 10^4$~\cite{ChlubaSunyaev2012,KhatriSunyaev2012,Chluba2013b}. However, since their amplitudes are predicted to be at least an order of magnitude below that of the fiducial $\mu$-distortion, we do not include them in our main analysis \cite{ChlubaJeong2014,Desjacques2015}.\footnote{In principle, we could perform a similar optimization analysis for an instrument configuration that would target these distortions, but in this paper, we focus on the $\mu$-distortion as the most pressing near-term target.} In Sec.~\ref{sec:conclusion}, we briefly comment on the feasibility of detecting these signals with \emph{SPECTER}.

Throughout this work we use the sky model and Fisher forecast set-up from A17. In the subsections below, we summarize the spectral distortion models. Details regarding the foreground signals are included in Appendix~\ref{sec:FG}.

\subsection{Blackbody component}
In addition to the $\mu$- and $y$-distortions, we need to fit for the deviation of the observed CMB blackbody spectrum from that at a reference CMB monopole temperature $T_{0}=2.7255$ K~\cite{Fixsen2009}. This is necessary because of the finite precision of the current constraints (i.e., the temperature of the CMB will be measured at higher precision by future experiments than it is currently known)~\cite{ChlubaJeong2014}. We use a first-order deviation characterized by a fractional temperature difference parameter, $\Delta_{\rm T}$, 
\begin{equation}
\begin{split}
    \Delta B_{\nu}=B_{\nu}(T_{\rm CMB})-B_{\nu}(T_{0})\approx\frac{\partial B_{\nu}{(T_{0})}}{\partial T} (T_{\rm CMB}-T_{0}) = I_{0}\frac{x^{4}e^{x}}{(e^{x}-1)^{2}}\Delta_{\rm T}\\
    \mathrm{and}\quad \Delta_{\rm T} \equiv (T_{\rm CMB}-T_{0})/T_{0},
\end{split}
\end{equation}
where $\nu$ denotes frequency, $I_{0}=\left(\frac{2h}{c^2}\right)\left(\frac{k_{B}T_{0}}{h}\right)^3 \approx 270$~MJy/sr, $x=\frac{h\nu}{k_{B}T_{0}}$ is the dimensionless frequency, $h$ is Planck's constant, $c$ is the speed of light, and $k_{B}$ is Boltzmann's constant. We use a fiducial value  $\Delta_{\rm T}=1.2\times10^{-4}$~\cite{Fixsen2009}.

\subsection{$\mu$-distortion}

The SED for the $\mu$-distortion is \cite{IllarionovSiuniaev1975, Zeldovich1969, SunyaevZeldovich1970, ChlubaKhatriSunyaev2012,Chluba2013a}

\begin{equation}
    \label{eq.mu}
    I_{\nu}^{\mu}=I_{0}\frac{x^{4}e^{x}}{(e^{x}-1)^{2}}\left[\frac{1}{\beta}-\frac{1}{x}\right]\mu
\end{equation}
where $\beta\approx 2.1923$ \cite{Chluba2013a}. The constant term in the parentheses accounts for the additional temperature shift to the reference blackbody spectrum at $T_{0}$ as a result of the energy injection and the value of $\beta$ ensures the conservation of photons \cite{Lucca2020,Lucca2023}.

For the amplitude of the $\mu$-distortion, we consider energy injection from dissipation of primordial density perturbations via photon diffusion --- a process known as Silk damping (\cite[e.g.,][]{Sunyaev1970,Daly1991,BarrowColes1991,Hu1994a,ChlubaKhatriSunyaev2012}), and the removal of energy from the photon bath due to adiabatic cooling of ordinary matter \cite{Chluba2005, ChlubaSunyaev2012}. The latter process results in a negative $\mu$-distortion that is roughly an order of magnitude smaller than the positive distortion sourced by damping, so the two contributions partially cancel~\cite{KhatriSunyaevChluba2012,Chluba2016}.  We adopt a fiducial value of $\mu =2 \times 10^{-8}$ based on current primordial power spectrum constraints, extrapolated to the small scales that source the $\mu$-distortion \cite{ChlubaKhatriSunyaev2012, Chluba2013b}. The $\mu$-distortion has a null frequency at $\approx 125$~GHz and a peak amplitude of $\approx 6$~Jy/sr (see Fig.~\ref{fig:fgs_sds}).

\subsection{$y$-distortion}
\begin{figure*}
    \centering
    \includegraphics[width=\textwidth]{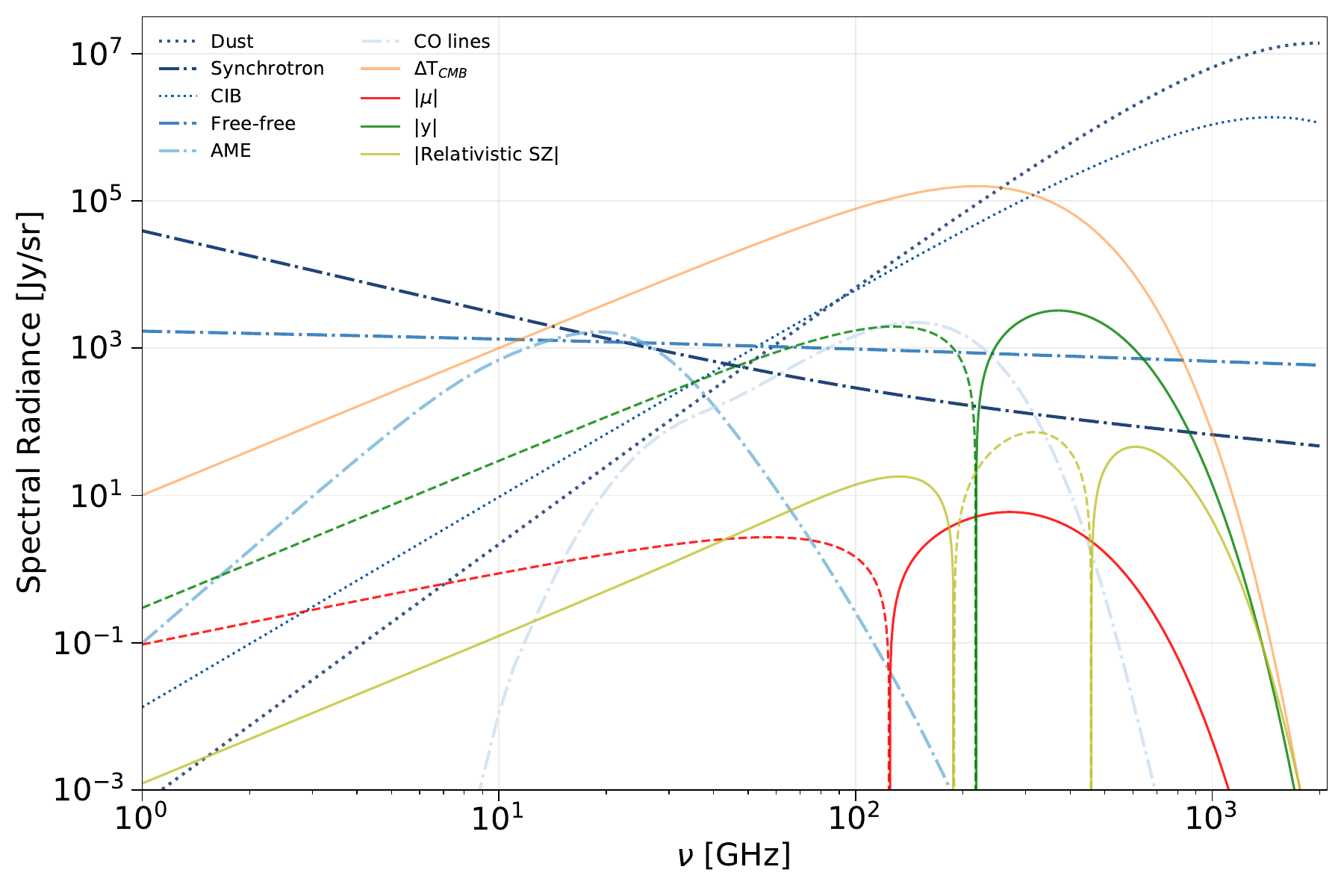}
    \caption{Sky signals used in our Fisher forecasts. Foregrounds are shown with blue curves. Galactic dust and CIB (dotted) dominate at high frequencies, while synchrotron (dot-dashed) dominates at low frequencies, with additional contributions from free-free, AME, and CO emission (dot-dashed). CMB spectral distortions are also plotted as labeled in the legend, with negative (positive) values indicated by dashed (solid) curves.}
    \label{fig:fgs_sds}
\end{figure*}

The $y$-distortion in the non-relativistic limit is given by \cite{Zeldovich1969,SunyaevZeldovich1970} 
\begin{equation}
    \label{eq.y}
    I^{y}_{\nu}=I_{0}\frac{x^4e^x}{(e^x-1)^2}\left[x\coth\left(\frac{x}{2}\right)-4\right]y
\end{equation}
where $y$ is the Compton-$y$ parameter, which is proportional to the integrated electron pressure along the line-of-sight. We use a fiducial value of $y=1.77\times10^{-6}$ based on the model of the tSZ monopole from Ref.~\cite{Hill2015}, and neglect any primordial contributions due to their much smaller amplitude ($\sim 2$-3 orders of magnitude lower~\cite{ChlubaKhatriSunyaev2012,Abitbol2017}), which in any case would be fit for in an actual data analysis, but indistinguishable from the late-time contribution. The tSZ signal is dominated by contributions due to CMB photons scattering off of electrons in the ICM, but also includes contributions from the IGM and the epoch of reionization. For a range of values based on observations and hydrodynamical simulations see Refs.~\cite{Chiang2020,Thiele2022}. The $y$-distortion has a null frequency at $\approx 218$ GHz and a peak amplitude of $\approx 3200$ Jy/sr (see Fig.~\ref{fig:fgs_sds}). 

Some ICM gas is heated to temperatures where relativistic effects are non-negligible. Using moments of the $y$-weighted ICM electron temperature distribution (following Refs.~\cite{Hill2015,Chluba_relSZ}), we can write these relativistic corrections to the tSZ monopole signal as follows (note that this contribution is added to that already written in Eq.~\eqref{eq.y}):
\begin{equation}
\begin{split}
    I^{\rm rel-SZ}_{\nu}=I_{0}\frac{x^4e^x}{(e^x-1)^2}\{ Y_{1}(x)\theta_{e}+Y_{2}(x)\theta_{e}^2+Y_{3}(x)\theta_{e}^3\}y.
\end{split}
\end{equation}
Here, $\Theta_{\rm e} \equiv k_{B}T_{eSZ}/m_{\rm e}c^{2}$, where $k_{B}T_{eSZ}$ is the first moment of the $y$-weighted ICM electron temperature distribution, for which we adopt a fiducial value of $k_{B}T_{eSZ}=1.245$ keV~\cite{Hill2015,Thiele2022}. The functions $Y_{i}(x)$ arise in the asymptotic expansion of the tSZ effect \cite{Sazanov1998,Challinor1998,Itoh1998,Nozawa2006}. The moment approach is necessary to properly capture the variations in ICM temperatures along the line of sight and across the sky (see Ref.~\cite{Hill2015} and Appendix A in Ref.~\cite{Thiele2022} for more details).\footnote{We explore the sensitivity of SPECTER to higher-order moments in Sec.~\ref{sec:conclusion}.} The relativistic tSZ SED has two null frequencies at $\approx 189$ GHz and $\approx 458$ GHz and a peak amplitude of $\approx 70$ Jy/sr (see Fig.~\ref{fig:fgs_sds}) .

\section{Fisher Forecast}\label{sec:fisher_forecast} 
We use a Fisher matrix formalism to evaluate \emph{SPECTER}'s capability to measure the $\mu$-distortion, while marginalizing over the other CMB signals (i.e., the blackbody, Compton-$y$, and relativistic tSZ contributions) described in the previous section and the foregrounds detailed in Appendix \ref{sec:FG}. We adopt the Fisher forecast set-up used in A17.\footnote{\url{https://github.com/asabyr/sd_foregrounds_optimize} (modified version of \url{https://github.com/mabitbol/sd_foregrounds})} As a validation of this approach, we note that Ref.~\cite{Sabyr2024} has recently carried out a re-analysis of the \emph{COBE/FIRAS} all-sky monopole data, and have shown that the error bars obtained from the re-analysis are in good agreement with the results of the Fisher forecast using this set-up.  In these calculations, we define the sky-averaged total monopole signal, $I_{\nu}$, after subtraction of an assumed blackbody SED at $T_{0} = 2.7255$~K, as:
\begin{equation}
\label{eq:totdistortion}
    I_{\nu}=\Delta B_{\nu} + I^{y}_{\nu}+I^{\rm rel-tSZ}_{\nu}+ I^{\mu}_{\nu}+  I_{\nu}^{\rm fg} .
\end{equation}
Here, $\Delta B_{\nu}$ is a blackbody component term, $I^{y}_{\nu}$ is the $y$-distortion, $I^{\rm rel-tSZ}_{\nu}$ is the relativistic contribution, $I^{\mu}_{\nu}$ is the $\mu$-distortion, and $I_{\nu}^{\rm fg}$ is the sum of all foreground contributions (see the equations in Sec.~\ref{sec:CMBSD} and Appendix~\ref{sec:FG}). In total our model has 16 free parameters. Their fiducial values and definitions are summarized below:\footnote{Additionally, in Appendices~\ref{app:emission} and ~\ref{app:zodiacal}, we study the impact of Galactic emission lines and the zodiacal light, respectively.}

\begin{enumerate}
    \item \textbf{Spectral distortions:}
    \begin{enumerate}
        \item \textbf{Blackbody temperature deviation ($\Delta B_{\nu}$):} $\Delta_{\rm T}=1.2\times10^{-4}$ -- fractional temperature difference (value chosen to lie within the constraints from Ref.~\cite{Fixsen2009}).
        \item \textbf{$y$-distortion ($I^{y}_{\nu}$):} $y=1.77\times10^{-6}$ -- monopole Compton-$y$ amplitude.
        \item \textbf{Relativistic correction to tSZ ($I^{\rm rel-tSZ}_{\nu}$)}: $k_{\rm B}T_{eSZ}=1.245$~keV -- first moment of the $y$-weighted electron temperature.
        \item \textbf{Primordial $\mu$-distortion ($I^{\mu}_{\nu}$):} $\mu=2\times10^{-8}$ -- monopole $\mu$-distortion amplitude.
    \end{enumerate}

    \item \textbf{Foregrounds ($I_{\nu}^{\rm fg}$):}
    \begin{enumerate}
        \item \textbf{Galactic thermal dust:} $A_{\rm D}=1.36\times10^{6}$~Jy/sr, $\beta_{\rm D}=1.53$, $T_{\rm D}=21$~K -- modified blackbody (MBB) SED amplitude, spectral index, and temperature, respectively.   
    \item \textbf{Galactic synchrotron:} $A_{\rm S}=288.0$~Jy/sr, $\alpha_{\rm S}=-0.82$, $\omega_{\rm S}=0.2$ -- power-law SED amplitude, spectral index, and logarithmic curvature index, respectively. 
    \item \textbf{Cosmic infrared background (CIB):} $A_{\rm CIB}=3.46\times10^{5}$~Jy/sr, $\beta_{\rm CIB}=0.86$, $T_{\rm CIB}=18.8$~K -- MBB SED amplitude, spectral index, and temperature, respectively.
    \item \textbf{Free-Free:} $A_{\rm FF}=300$~Jy/sr --  amplitude of the model from Ref.~\cite{Draine2011}.
    \item \textbf{Integrated CO:} $A_{\rm CO}=1$ -- dimensionless amplitude for the template, which was computed using the spectra from Ref.~\cite{Mashian2016}. 
    \item \textbf{Spinning dust:} $A_{\rm AME}=1$ -- dimensionless amplitude for the template, which was computed using the model in Ref.~\cite{Planck2016FG}.
    \end{enumerate}
\end{enumerate}

Using these distortion and foreground models, we calculate the Fisher information matrix as 
\begin{equation}
    F_{ij}= \sum_{\nu,\nu'}\frac{\partial I_{\nu}}{\partial p_{i}}C_{\nu\nu'}^{-1}\frac{\partial I_{\nu}}{\partial p_{j}} \,,
\end{equation}
where $\nu,\nu'$ are the central frequencies of the instrument passbands; $p$ corresponds to distortion and foreground parameters that we vary and is indexed with $i,j$; and $C_{\nu\nu'}$ is the \emph{SPECTER} noise covariance matrix, which we assume to be diagonal. We do not explicitly integrate the sky signal SEDs over the passbands of the instrument but instead just evaluate them at the central frequencies, for computational efficiency. This was shown to be a small effect in A17 for \emph{PIXIE}. The parameter covariance matrix is determined by inverting $F_{ij}$.
We use this formalism to determine the optimal bands and detector counts for a $\mu$-distortion measurement in Sec.~\ref{sec:configuration_optimization}. Unlike the fiducial set-up in A17, we do not impose external priors on any of the parameters, instead determining all of them with \emph{SPECTER}.

%%%%%%%%%%%%%%%%%%%%%%%%%%%%%%%%%%%%%%%%%%%%%%%%%%%%%%%

\section{Instrument Sensitivity Forecast}\label{sec:bolocalc}
To forecast the instantaneous sensitivity of \emph{SPECTER}, we divide the instrument into three discrete frequency ranges utilizing different technologies that are optimized for each spectral range. At low frequencies ($\nu < 10$~GHz), we assume high electron mobility transistor (HEMT) amplifier-based differential radiometers. At intermediate frequencies ($10 < \nu < 500$~GHz), we assume bolometric detectors with on-chip bandpass filtering made of superconducting niobium microstrip. At high frequencies ($\nu>500$~GHz) beyond the superconducting gap of niobium, we baseline direct absorption bolometers with free-space filtering. We build a detailed sensitivity model that encompasses the differences between these technologies, as well as the expected sky emission outlined in Appendix \ref{sec:FG} and various other experimental details. In this section, we briefly discuss the specifics of these sensitivity calculations, while leaving further instrument details for Sec. \ref{sec:SPECTER}. Instantaneous sensitivity values are given per individual detector: a single detector diode for low frequencies or a single bolometer for higher frequencies.

\paragraph{Bolometric Channels:}
To obtain sensitivity estimates for the bolometric channels, we use the \texttt{BoloCalc} sensitivity calculator~\cite{Hill2018} to build a detailed sensitivity model of the instrument. This model uses a broad range of low-level detector and optical parameter inputs to produce a white-noise sensitivity for each spectral channel evaluated in the Fisher forecast. The noise estimate, represented as a noise-equivalent temperature (NET) in CMB temperature units, includes contributions from photon noise, thermal bolometer noise, and electrical readout noise. The photon noise budget is driven by the total optical power incident on the detectors, $P_\textrm{opt}$, which includes both astrophysical and instrumental thermal emission. For each spectral channel, this can be expressed as

\begin{equation}
    \label{eq:p_opt}
    P_\textrm{opt} = \int^\infty_0 \left( \sum^N_i p_i(\nu) \right) f(\nu) d\nu \,
\end{equation}
\noindent where $f(\nu)$ is the passband response of the instrument channel, $N$ is the total number of emissive blackbody elements—including the CMB itself—between the detector and the sky, and $p_i(\nu)$ is the power spectral density of the $i$th emissive element in the chain. Note that $p_i(\nu)$ will depend on the temperature and emissivity of element $i$, as well as the total optical coupling efficiency between the element and the detector. For foreground emission, we use the sky model described in Appendix \ref{sec:FG}, integrated across the bandwidth of each spectral channel.

The bolometer and readout noise budgets are largely driven by detector saturation powers and operating temperatures. For \emph{SPECTER}, we assume a saturation power equivalent to three times the expected loading, $P_{\rm opt}$, and a 100~mK operating temperature. For all other detector and readout parameters, we use estimates informed by existing technologies that have been demonstrated on previous experiments. 

\paragraph{Low-Frequency Channels:}
At frequencies below approximately 10 GHz, the optical loading from the sky decreases significantly and fabricating bolometers with sufficiently low saturation powers becomes a major challenge. Coherent radiometers are thus better suited for observing at these low frequencies. For \emph{SPECTER}, we assume a \emph{WMAP}-style pseudo-correlation differential radiometer design~\cite{Jarosik_2003} using HEMT amplifiers as the baseline. We calculate the sensitivities of the HEMT-based spectral channels separately from the bolometric projections provided by \texttt{BoloCalc}. For a differential radiometer with bandwidth $\Delta\nu$ and optical efficiency $\eta$, the instantaneous sensitivity of a single detector diode is given as 
\begin{equation}\label{eq:LFFI_delta_T}
    \mathrm{NET} = 2\frac{T_\textrm{sys}}{\sqrt{\eta \Delta \nu}} \ .
\end{equation}
% $\textrm{NET} = T_\textrm{sys}\sqrt{2}/\sqrt{\eta \Delta \nu}$\footnote{Note that although there is an additional factor of $\sqrt2$ due to swapping between the sky and the calibrator source, this degradation in sensitivity is already taken into account in Eq.~\ref{eq:noise} since we always only consider $t_{\rm obs}$, the time spent observing the sky, in the forecasts.}
In this design, each feedhorn contains two detector diodes per polarization channel for a total detector count of four. We assume that amplifier noise dominates the system temperature $T_\mathrm{sys}$, with the factor of two accounting for the inherent differencing in this design as described in Section \ref{sec:SPECTER}. For all HEMT channels, we set $\eta=0.35$, consistent with the end-to-end efficiencies calculated for the bolometric channels. To estimate the amplifier noise, we use the temperature values provided for commercially-available HEMT amplifiers from Low Noise Factory\footnote{\url{https://lownoisefactory.com/}}. Finally, we note that the sky-calibrator swapping design of \emph{SPECTER} ultimately implies an additional $\sqrt2$ sensitivity degradation to all channels due to loss in observing time. However, since we always use $t_{\rm obs}$, the time spent observing the sky, when computing the integrated sensitivity (Eq.~\ref{eq:noise}), we do not need to include this additional factor in Eq. ~\ref{eq:LFFI_delta_T}.

\paragraph{Sensitivity Validation:}
The predictive power of these noise forecasts are validated using a similarly detailed model of the \emph{Planck} High-Frequency Instrument (HFI), for which realized and well-understood performance values~\cite{Planck2011} can be used for comparison. The \emph{Planck} HFI instrument model is derived from publicly-available parameters that define the passband, emissivity, and optical efficiency of its single-moded instruments at 100, 143, 217, and 353~GHz. The simulated NET values agree with on-sky measurements~\cite{Planck2011} to within about 10\%. The higher-frequency channels at 545 and 857~GHz cannot be validated in this manner due to the complex nature of their multi-moded optics. However, we note that only single-moded detectors are considered in the design of \emph{SPECTER}.

%%%%%%%%%%%%%%%%%%%%%%%%%%%%%%%%%%%%%%%%%%%%%%%%%%%%%%%

\section{Instrument Configuration Optimization}\label{sec:configuration_optimization}

With \emph{SPECTER}'s instrument design, we have flexibility in setting the exact frequency bands and their individual sensitivities --- i.e., we can choose the overall frequency range, the number of bands and their respective widths, and the number of detectors for each band. In principle, there could exist several possible set-ups that are both feasible to build and near-optimal for observing the $\mu$-distortion. We note that our optimization results are predicated on our assumed fiducial sky model, but we assess the impact of this choice in detail in Sec.~\ref{sec:sky_modeling_choices}. 

To find one viable instrument configuration, we use two quantities to assess our instrument set-up: the forecast signal-to-noise ratio (SNR) of the $\mu$-distortion signal and the approximate focal plane area of the instrument. The latter quantifies the relative cost of the instrument, since the limiting factor for a satellite design is its physical size, which is primarily driven by the total size of all the detectors.\footnote{We therefore will use focal plane area, area, and cost interchangeably throughout this paper.}

The SNR for a given parameter $p_{i}$ can be computed via the Fisher matrix formalism (described in Sec.~\ref{sec:fisher_forecast}) and is a function of the instrument's sensitivity at each observed frequency or diagonal noise covariance matrix, $C_{\nu\nu}$:
\begin{equation}
    \mathrm{SNR}_{p_{i}}(C_{\nu\nu})=\frac{p_{i, \rm fid}}{\sigma_{p_{i}}(C_{\nu\nu})}=\frac{p_{i, \rm fid}}{\sqrt{(F^{-1})_{ii}(C_{\nu\nu})}} \,.
\end{equation}
Since we always use a diagonal noise covariance in this work, we will use $C_\nu$ from now onwards to refer to the \emph{SPECTER} noise at frequency $\nu$. To compute $C_{\nu}$, we first follow the procedure outlined in Sec.~\ref{sec:bolocalc} to calculate the instantaneous sensitivity, or NET, for a given spectral channel. We then convert the per-detector NET from $\mu {\rm K}_{\rm CMB} \sqrt{\rm s}$ to Jy/sr (the units used in the Fisher calculations) assuming observation time, $t_{\rm obs}$, and the number of detectors at each frequency channel, $N^\mathrm{det}_{\nu}$, as follows:

\begin{equation}\label{eq:noise}
\begin{split}
\left(\frac{C_{\nu}}{\mathrm{Jy/sr}}\right)=\left(\frac{\mathrm{NET}}{\mu \mathrm{K_{CMB}\sqrt{\rm s}}}\right)\left(\frac{1}{\sqrt{\mathrm{t_{\rm obs}/s}}}\right)
\left(\frac{1}{\sqrt{N^{\rm det}_{\nu}}}\right)\\
\times \left(\frac{\partial B_{\nu}}{\partial T}\bigg\rvert_{T=T_0}\frac{\mu \mathrm{K_{CMB}}}{\mathrm{Jy/sr}}\right).
\end{split}
\end{equation}
Here, $\frac{\partial B_{\nu}}{\partial T}\big\rvert_{T=T_0}$ is the conversion factor between CMB thermodynamic temperature units and specific intensity units (see more details in Appendix~\ref{app:units}). Due to the need to swap between the sky and the blackbody source for absolute calibration, $t_{\rm obs}$ corresponds to the time spent observing the sky, which is half of the total mission duration time. The other half of the time is spent observing the internal calibrator (see Sec. \ref{sec:SPECTER} for more details). Note that $N^{\rm det}$ corresponds to one detector diode and one bolometer for HEMT and bolometer frequencies, respectively, as described earlier in Section ~\ref{sec:bolocalc}.

For the detector technology in the \verb|BoloCalc| sensitivity calculations, we always assume HEMT amplifiers for $\nu$ < 10 GHz and bolometers for $\nu$ > 10 GHz. Bolometers formally have slightly better sensitivities than HEMT amplifiers, but we cannot use bolometers at $\nu$ < 10 GHz due to the technical limitations described in the previous section, so we choose to use them to the lowest possible frequency, $\nu=10$~GHz (see Sec. \ref{sec:bolocalc}). We use the focal plane area as a proxy for the cost of the instrument. To estimate the total area, we assume that a single detector centered at $\nu=150$~GHz is coupled to a feedhorn antenna occupying a focal plane area of $5.25^{2}$~mm$^{2}$~\cite{Simon2016} and scale the areas of all other detector channels as $1/\nu^{2}$. The area, $A$, is then a function of \emph{SPECTER}'s central frequencies and is summed over both the frequencies and the number of detectors: 

\begin{equation}
\begin{split}
A=\sum_{\nu}\frac{150^{2}\,\,\mathrm{GHz}\times5.25^{2}\,\,\mathrm{mm}^{2}}{\nu^{2}}\times N_{\nu}^{\rm det}\times \mathrm{DF}_{\nu}\\
    \mathrm{where} \quad \mathrm{DF}_{\nu} = \begin{cases} 
    \frac{1}{4} & \nu<10 \,\mathrm{GHz}\\
      \frac{1}{2} & 10 \,\mathrm{GHz} \leq\nu<500 \,\mathrm{GHz}\\
      1 & \nu\geq500 \, \mathrm{GHz} \,.
   \end{cases}
\end{split}
\end{equation}
Here, DF stands for detectors per feedhorn. We assume that feedhorns at $\nu<500$~GHz are dual-polarization-sensitive. Effectively, at these frequencies we have two detectors for the cost of one (i.e., two detectors per feedhorn antenna). For HEMT frequencies, we have two detector diodes per polarization resulting in four detectors per feedhorn in total. 

Although we have both a clear cost function and a quantity that we want to maximize, we do not use a standard constrained optimization approach (e.g., via Lagrange multipliers). This is because the $\chi^{2}$ (cost) surface in our case is expected to be very uneven, with valleys of local minima. Therefore, to systematize our approach in finding a possible instrument configuration, we follow two main steps: band and detector count optimizations. We first select the bands that are optimal for measuring the $\mu$-distortion, as described further below. Then we determine the number of detectors needed at each band to reach a $\sim 5\sigma$ detection for a specified integration time and sky fraction used in the final analysis, while keeping the overall area of the instrument below a specified threshold.

\subsection{Frequency Band Optimization}
Our band selection process can be summarized in a few simple steps:
\begin{enumerate}
    \item Start with a large number of narrow bands ($N_{\rm bands}$), specified further below, covering some frequency range $\nu\in \{\nu_{\rm min}, \nu_{\rm max}\}$ and compute the SNR$_{\mu}$ assuming a single detector per band. 
    \item Combine a pair of neighboring bands into a wider band and check the effect on the SNR$_{\mu}$ with this new configuration. Repeat this process for each pair. After checking every neighboring pair, combine the pair that has the most optimal effect on the SNR$_{\mu}$ --- i.e., the combination that results in either the smallest decrease or the largest increase in the SNR$_{\mu}$. After one full iteration over frequencies, we are left with $N_{\rm bands}-1$ bands. 
    \item Repeat step (2) until SNR$_{\mu}$ drops to a chosen fraction of the starting SNR$_{\mu}$ (specified in detail below).
\end{enumerate}

\indent The band selection process is driven by the trade-offs between resolution and detector noise. Wider bands have lower noise, but spectral resolution is needed to map the sky signals across different frequencies at sufficient precision. We start our band selection process with many narrow bands and corresponding initial SNR$_{\mu}$. As we combine the neighboring pairs into wider bands, the detector noise decreases resulting in an increase in SNR$_{\mu}$. At some point in this iterative process, we reach peak sensitivity, after which, due to a substantial decrease in spectral resolution with fewer bands, the sensitivity to the $\mu$-distortion signal starts decreasing (because the foreground marginalization noise penalty becomes large).

To find an optimized frequency configuration for detecting the $\mu$-distortion, we need to define an initial set of narrow bands $\nu_{i}$, between some minimum, $\nu_{\rm min}$, and a maximum band edge frequency, $\nu_{\rm max}$. To help set the initial bands, we use the following restrictions --- (1) bands cannot be narrower than 1 GHz ($\Delta\nu \nless 1$ GHz), (2) at 10 GHz, $\Delta\nu/\nu\sim0.3$, and (3) the minimum allowed $\Delta\nu$ increases with frequency. This choice does not exactly reflect the physical limitations for different frequency ranges, but is a convenient starting point to have sufficiently narrow bands to allow flexibility in the final configurations. We use constraints (1) and (2) to set the minimum $\Delta\nu_{0}^{\rm H}=1$ GHz for the HEMT (H) amplifiers and $\Delta\nu_{0}^{\rm B}=3$ GHz for the bolometers (B). To decrease the total number of initial bands/iterations, motivated by the constraint (3), we increase $\Delta\nu^{\rm B}$ for every frequency interval of $100$ GHz (we are only using HEMTs up to 10 GHz so this is only relevant to bolometers). Since the final bands will be wider than the initial ones and reflect some combination of the neighboring pairs, we do not expect the exact choice of the initial bands to affect our results significantly. The setting of the initial bands can be summarized as follows,
\begin{equation}
\begin{split}
    \nu_{i}=\{(\nu_{n, a},\nu_{n, b}),(\nu_{n+1, a},\nu_{n+1, b}), ...\}\\
    \mathrm{where}\quad \mathrm{e.g.}\quad\nu_{n, a}\equiv\nu_{\rm min}, \quad \nu_{n, b}=\nu_{n, a}+\Delta\nu_{0}, \quad \nu_{n+1, a} \equiv \nu_{n, b}\\
\end{split}
\end{equation}
Here, $n$ corresponds to the band number in the sequence, $a$ ($b$) denote the lower (higher) frequency edges of a given band, and the bands are more generally set as
\begin{equation}
    \begin{split}
        \nu_{n+1, b}=\rm{min}\{\nu_{n+1, a}+\Delta\nu_{n},\nu_{\rm max}\}\\
    \mathrm{where}\quad \Delta\nu_{n}=\Delta\nu_{0}+\floor*{\frac{\nu_{n+1, a}}{100\,\mathrm{GHz}}}\Delta\nu_{0} 
    \end{split}
\end{equation}

\noindent We take $\nu_{\rm min}^{\rm H}=1$ GHz,  $\nu_{\rm max}^{\rm H}=10$ GHz, $\Delta\nu_{0}^{\rm H}=1$ GHz for the HEMT amplifiers, and
$\nu_{\rm min}^{\rm B}=10$ GHz,  $\nu_{\rm max}^{\rm B}=2000$ GHz, $\Delta\nu_{0}^{\rm B}=3$ GHz for the bolometers so $\nu_{i}=\{\nu_{i}^{\rm H},\nu_{i}^{\rm B}\}$.

Using these initial bands, we perform the iterative band optimization process described above. In this part of the configuration search, we fix  $t_{\rm obs} = 6$ months, and $N_{\nu}^{\rm det, B}=\bf{1_{\nu}}$, a vector of ones for bolometers, and for the HEMT frequencies, we assume sensitivities with $N_{\nu}^{\rm det, H}=\bf{4_{\nu}}$.  We perform this iterative selection process for a set of different minimum edge frequencies $\nu_{\rm min}^{\rm H}=\{1, 2, 3, 4, 5\}$ GHz with $\nu_{\rm min}^{\rm B}=10$ GHz, as well as using only the bolometers (i.e., $\nu_{i}=\{\nu_{i}^{\rm B}\}$) with $\nu_{\rm min}^{\rm B}=\{10, 30\}$ GHz. For each of these set-ups, we check three final frequency configurations: (1) the final sensitivity and frequency configuration before dropping to $25\%$ of the initial SNR$_{\mu}$, (2) $50\%$ of the initial SNR$_{\mu}$, and (3) at its peak value before it starts to decrease due to the small number of bands. For each of those minimum frequency choices, we select the result that has the highest ratio of SNR$_{\mu}$ to area $A$ from options (1), (2), and (3).

In addition to these frequency configurations, we also check starting with $\Delta\nu_{0}^{\rm H}=\{2,3,4\}$ GHz for the HEMT amplifiers but find that using the narrowest width gives the best results. Similarly, we additionally check switching to bolometer technology only at 30 GHz (using $\Delta\nu^{\rm H}_{0}=1$ GHz now up to $\nu_{\rm max}^{\rm H}=30$ GHz and $\Delta\nu^{\rm B}_{0}=1$ GHz with $\nu_{\rm min}^{\rm B}=30$ GHz), but find that we get better results by switching to bolometers at 10 GHz. From these options, we select the frequency set-ups that can reach SNR$_{\mu}$ $\gtrsim 5$ with 200 (100) detectors per band for HEMTs (bolometers), still assuming $t_{\rm obs}=6$ months, which leaves us with the frequency band option that has a minimum frequency of 1 GHz ($\nu_{\rm min}^{\rm H}=1$ GHz). As a reference, we started with 127 bands and ended up with a 16-band configuration. The exact band edges and central frequencies are listed in Table~\ref{tab:fid_setup} in Columns 1 and 2, respectively. Their instantaneous per-detector noise in temperature units is listed in Column 3.

\begin{figure}
    \centering
    \includegraphics[width=0.6\textwidth]{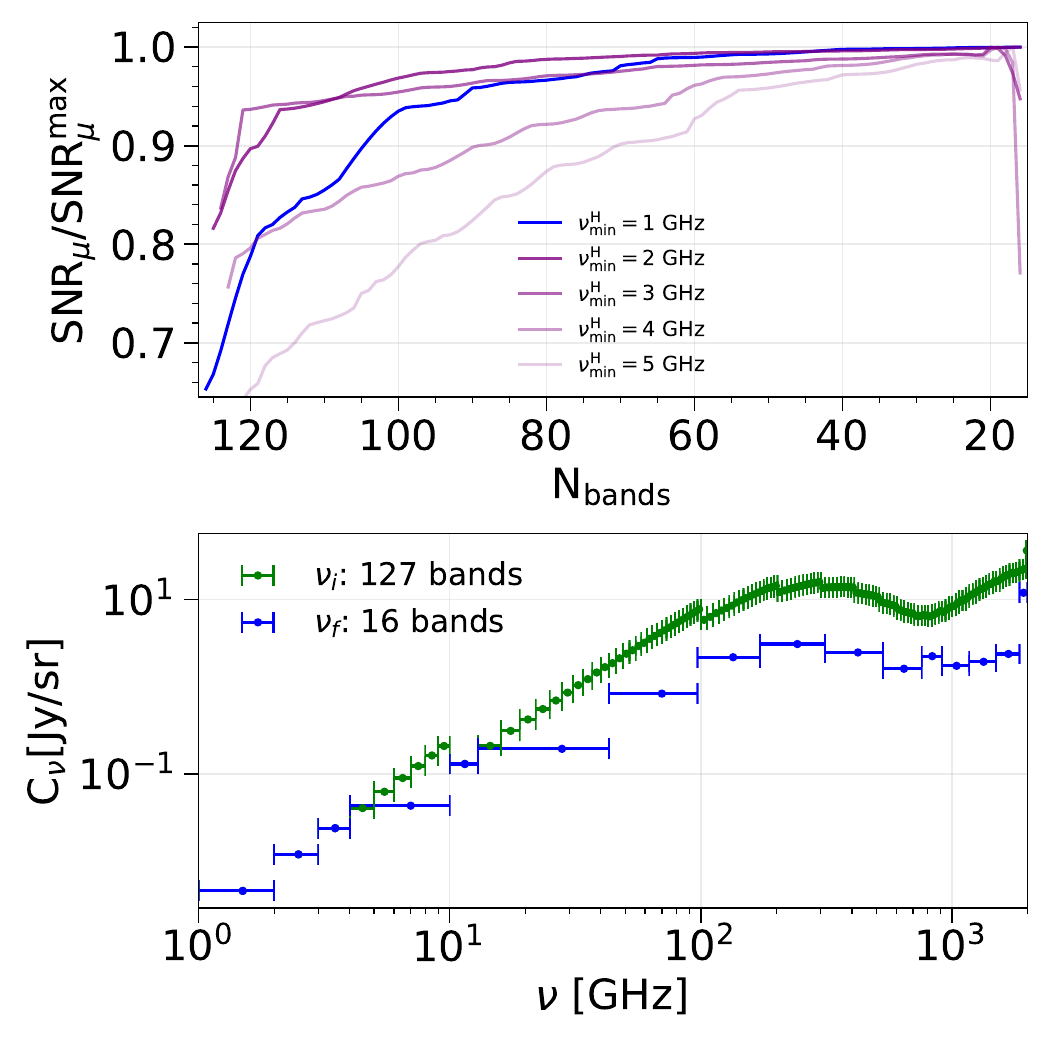}
    \caption{Visualization of the band optimization process. The SNR$_{\mu}$ and sensitivities in this plot are computed assuming 1 detector per band for bolometers and 4 detector per band for HEMT amplifiers and $t_{\rm obs}=6$ months. \textit{Top:} The SNR$_{\mu}$ at each iteration step in the band-optimization process for several example set-ups using different minimum frequencies. We start with a set of narrow initial bands and at each step combine a pair that will have the most optimal effect on SNR$_{\mu}$.  The sensitivity increases with increasing bandwidth, but once the frequency resolution is too low, the SNR$_{\mu}$ drops. \textit{Bottom:} Comparison between the initial bands $\nu_{i}$ and the optimized configuration, $\nu_{f}$. Note the scale on the y-axis: the noise is many orders of magnitude larger at the highest frequencies due to the photon loading from the high-frequency foregrounds.
    }
    \label{fig:band_optimization}
\end{figure}

In Fig.~\ref{fig:band_optimization}, we show a visualization of the band-optimization process for the configuration with $\nu_{\rm min}^{\rm H}=1$ GHz and several other examples with different $\nu_{\rm min}^{\rm H}$. The upper panel in the plot shows SNR$_{\mu}$ as a function of the number of bands at each step. The first point corresponds to the initial  SNR$_{\mu}$ using a set of initial narrow bands as described earlier. The rest of the points correspond to SNR$_{\mu}$ at each step of the optimization process when an optimal pair of neighboring bands is combined. SNR$_{\mu}$ steadily increases then drops once the frequency resolution becomes too low. In the case of the optimized configuration that we adopt ($\nu_{\rm min}^{\rm H}=1$), the peak SNR$_{\mu}$ is reached just before the SNR$_{\mu}$ drops significantly and therefore options (1), (2), and (3) correspond to the same frequency configuration. For other examples shown in the figure, a drop in SNR$_{\mu}$ can be clearly seen. The bottom panel also shows the initial set of narrow bands versus the optimized bands for our chosen configuration. Note the large difference in scale in the sensitivities at low and high frequencies. The noise is larger at the highest frequencies due to foreground loading from the dust emission. 

\subsection{Detector Count Optimization}
For the detector count optimization, we perform a grid-search computation over the possible number of detectors for each band. Since the number of computations increases rapidly with the number of bands or the grid dimension, it is time-prohibitive to perform a fully exhaustive search --- varying each of the bands for every value of detector counts in a feasible range simultaneously --- even for just the 16-band configurations. Thus, we instead perform a series of coarser grid computations and narrow down to a possible configuration that achieves the desired SNR$_{\mu}$, while minimizing the cost. 

In our first set of detector count calculations, we vary the number of detectors for bands in groups of 4 (i.e., 16 bands are split into 4 groups and each group has the same number of detectors), allowing each frequency group to vary the number of detectors between 2-200 for HEMT amplifiers (lowest group of 4 bands) and 1-100 for bolometers with 50 values each (i.e., a grid of $50^{4}$ calculations). This initial calculation is motivated by multichroic technology --- feedhorn pixels can be shared between neighboring bands. This is physically motivated since some neighboring bands may realistically share the same feedhorn on the instrument and allows us to perform faster evaluations by grouping bands into pairs or groups and thus reducing the size of the grid dimensions. From this test, we find that the configuration that extends to 1 GHz is able to achieve a $5\sigma$ measurement with the lowest-frequency bands having of order $\sim 30$ detectors. As a comparison, if we perform the initial grid computations for the configuration that starts with 2 GHz, we find that SNR$_{\mu}$ can only reach SNR$_{\mu}$ = 2.4, while maintaining the same area as the configuration that goes down to 1 GHz and achieves $\sim5\sigma$. 
% Since we can only realistically use $<20$ detectors per band at the lowest frequencies due to their size, we choose to take the 1-2000 GHz configuration as our optimized choice. 
% The exact band edges and central frequencies are listed in Table~\ref{tab:fid_setup} in Columns 1 and 2, respectively. Their instantaneous per-detector noise in temperature units is listed in Column 3.

\begin{table*}
  \begin{threeparttable}[t]
  \cprotect\caption{\label{tab:fid_setup} Sensitivities for the fiducial \emph{SPECTER} set-up. The table lists band edges [GHz], central frequencies [GHz], instantanous per-detector noise [$\mu{\rm K}_{\rm CMB}\sqrt{s}$], number of detectors, number of detectors per feedhorn, integrated noise [Jy/sr], and map-domain noise [$\mu {\rm K}_{\rm CMB}-$arcmin] assuming $t_{\rm obs}=1$ year and a full-sky observation. The expected angular resolution ranges from $\sim1$ degree at 150~GHz to $\sim 5$ degrees at 1.5~GHz. The total number of detectors is $N^\mathrm{det}=1100$.}
   \centering
   \begin{tabular}{|l|c|c|c|c|c|c|}
     \hline
Band & $\nu$& NET & $N^\mathrm{det}_{\nu}$ & Detectors &$C_{\nu}$ & $C_{\nu}^{\rm map}$\\
$\left[ {\rm GHz} \right]$ & $\left[ {\rm GHz} \right]$ & $\left[ \mu{\rm K}_{\rm CMB}\sqrt{\rm s} \right]$ & &per feedhorn& $\left[ 10^{-2} \,\, {\rm Jy/sr}\right]$& $\left[\mu{\rm K}_{\rm CMB}-\mathrm{arcmin}\right]$\\
% %\colrule
\hline
1-2&1.5&518.5&8&4&0.2256&397.7\\
2-3&2.5&490.6&16&4&0.4191&266.0\\
3-4&3.5&500.5&24&4&0.6843&221.6\\
4-10&7&227.1&24&4&1.241&100.6\\
10-13\tnote{*}&11.5&127.1&8&2&3.240&97.51\\
13-43&28&32.53&20&2&3.057&15.78\\
43-97&70&24.91&100&2&5.890&5.403\\
97-172&134.5&24.37&100&2&15.42&5.287\\
172-313&242.5&25.91&100& 2&21.85&5.620\\
313-532&422.5&55.07&100&2&17.50&11.95\\
532-757&644.5&329.7&100&1&11.37&71.51\\
757-913&835&4671&100&1&15.85&1013\\
913-1174&1043.5&$5.836\times10^{4}$&100&1&12.29&$1.266\times10^{4}$\\
1174-1501&1337.5&$4.259\times 10^{6}$&100&1&13.66&$9.240\times10^{5}$\\
1501-1858&1679.5&$8.699\times 10^{8}$&100&1&16.82&$1.887\times10^{8}$\\
1858-2000&1929&$2.040\times 10^{11}$&100&1&84.85&$4.426\times10^{10}$\\
%\colrule
\hline
\end{tabular}
     \begin{tablenotes}
       \item [*] \footnotesize{Note that in the final instrument design shown in Fig.~\ref{fig:focal-plane}, we choose to use a HEMT amplifier for this band, with 40 detectors as described in Sec.~\ref{sec:SPECTER}. This has a $\sim29\%$ effect on the band sensitivity, $C_{\nu}$, but $\lesssim1\%$ effect on the forecast SNR$_{\mu}$. Therefore, we leave all the forecast values in the paper as they were computed with the original choice of bolometer technology for this band. The same detector technology change applies to the set-up in Table~\ref{tab:multichroic}}.
     \end{tablenotes}
  \end{threeparttable}
\end{table*}

\begin{figure}
    \centering
    \includegraphics[width=0.6\textwidth]{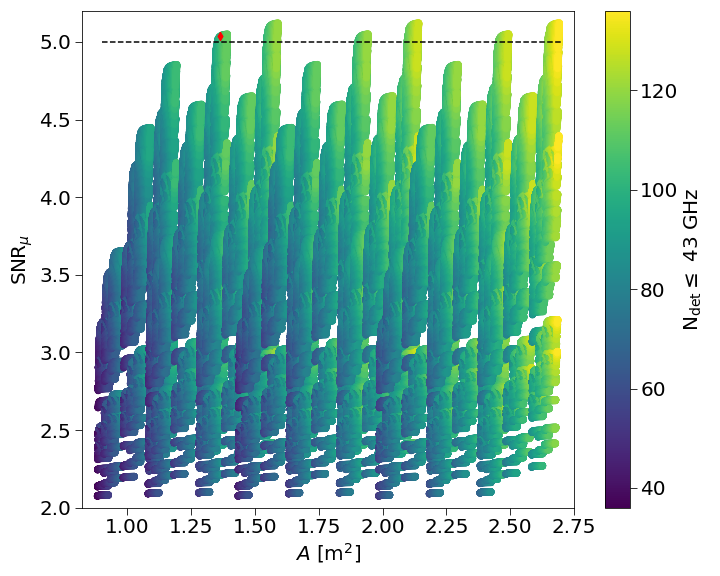}
    \caption{Visualization of the detector optimization: SNR$_{\mu}$ versus focal plane area for a grid of 5,062,500 instrument configurations. The points are colored by the sum of the number of detectors in the six lowest-frequency bands ($<43$ GHz, see Table \ref{tab:fid_setup}), since the low-frequency detectors are the largest in area and therefore drive the size and cost of the instrument. The dotted line marks SNR$_{\mu}= 5$.  Our chosen configuration is marked with a diamond marker (red).}
    \label{fig:det_optimization}
\end{figure}

Based on the physical detector count limits at low frequencies, we select our final set-up using a grid computation where we allow the first 4 bands to vary between 8-24 detectors, 5-6th bands to vary between 2-20  detectors, and 7-8th bands to vary between 2-100 detectors, individually, and the highest 8 bands to vary in groups of 4 between 20-100 detectors (i.e., a grid of $3\times3\times3\times3\times10\times10\times5\times5\times5\times5=5,062,500$ calculations). Using a realistic nominal integration time of $t_{\rm obs}=12$ months and assuming $f_{\rm sky}=0.7$ (i.e., dividing $C_{\nu}$ by $1/\sqrt{f_{\rm sky}}$), we choose one of the configurations that reaches SNR$_{\mu}$ = 5, while keeping the area relatively small at $A\sim1.36$ m$^{2}$ as shown in Fig.~\ref{fig:det_optimization}.
% We perform the latter step to avoid too much fine-tuning in detector numbers, especially at high frequencies, where the detectors are small and the overall area is therefore roughly the same.

In Fig. \ref{fig:det_optimization}, we show the results from the final grid calculation used to select the optimized configuration. Every point corresponds to a Fisher forecast computation for a given detector count configuration. Our chosen configuration (red diamond marker) is one of many possible choices depending on the exact SNR$_{\mu}$ and $A$ cutoffs. Note that the area could be further decreased if we relaxed the SNR$_{\mu}$ requirement to, e.g., 4. We color the points by the sum of the number of detectors in the lowest six bands ($<43$ GHz) to further illustrate the relative cost for each configuration, since low-frequency detectors most strongly drive the cost of the instrument (i.e., they are the largest in size). We see clearly that the points that are the most sparse and require the highest number of low-frequency detectors are at SNR$_{\mu} \gtrsim 4.5$. The optimized detector counts and associated sensitivity, assuming a 1-year, full-sky observation, in Jy/sr are given in Table~\ref{tab:fid_setup} in Columns 4 and 6, respectively. We list the number of detectors allowed per feedhorn in Column 5.\footnote{The size of the final focal plane is $\approx1.6$~m$^{2}$ due to added space between detector arrays.}

Since the detectors at high frequencies have smaller size, we check whether increasing the number of detectors in the highest-frequency bands can reduce the number of detectors required at low frequencies. We find that increasing the sensitivity at higher frequencies does not sufficiently compensate and we still need a similar number of detectors at low frequencies to obtain the desired SNR$_{\mu}$. As an additional comparison, we run the same final grid calculation for the configuration that starts with 2 GHz and find that the highest SNR$_{\mu}$ that can be reached is $\simeq 1.6$.

In addition to the 16-band optimized set-up, we also consider a configuration that takes advantage of on-chip multichroic technology that has been demonstrated at intermediate frequencies, i.e., a single feedhorn sensitive to several narrower bands via multiplexed detectors~\cite{McMahon_2012}. To implement this, we split all octave bandwidths between 10 and 500~GHz into four sub-bands, thereby increasing frequency resolution at the expense of some sensitivity with each split (for the 6 bands that we split, the sensitivity is reduced between $\approx 5-18\%$). The motivation for such a choice is to improve the instrument's robustness to differences between the true foregrounds and those assumed in the model we have used for optimization, via increased spectral resolution. These additional band splittings give us a configuration with 34 bands, which results in SNR$_{\mu}=4.5$ for $t_{\rm obs}=1$ year. Note that the number of detectors and the detector array area remain the same as in the 16-band case via the multichroic technology. For $t_{\rm obs}=4$ years, the optimized 16-band configuration and this set-up, which we will refer to as ``34-band multichroic'', give SNR$_{\mu}=10$ and $9$, respectively. The 34-band multichroic set-up is characterized in Table~\ref{tab:multichroic}.\footnote{We take full advantage of the multichroic detectors for our 34-band multichroic example set-up, but this is not required. For example, one could consider a 28-band configuration, which uses multichroic technology between 40-500 GHz, which gives slightly better SNR$_{\mu}=4.9$ and SNR$_{\mu}=9.9$ respectively for the fiducial and extended duration, but has somewhat coarser frequency resolution.}

Since the cost is primarily driven by the lowest-frequency channel (due to its large detector size), we experiment with dropping this channel from the configuration. 
% As noted earlier, we find during optimization that the number of detectors required for a configuration that extends to 2 GHz is larger than that required for a configuration that extends to 1 GHz. 
If we drop the lowest frequency in the 34-band multichroic set-up\footnote{In the 16-band case, there are too few frequency channels to fit for all of the foregrounds, so dropping the lowest one results in an unstable SNR$_{\mu}$.}, we find that SNR$_{\mu}$ drops from $4.5$ to $0.8$ and from $9$ to $1.6$ for 1- and 4-year observation times, respectively. Placing $10\%$ external priors on the synchrotron spectral index and amplitude does not substantially improve the results. We therefore conclude that it is very important for the instrument's frequency coverage to extend to 1~GHz to maintain a sufficiently low number of detectors in our set-up.

As previously stated, we compute all forecasts in this work by evaluating our sky model at the central frequencies of each channel, because it was shown that integrating over passbands did not have a large effect for \emph{PIXIE} forecasts in A17. Since some of the \emph{SPECTER} bands are wider than those considered in A17, we check the impact of passband integration on our optimized \emph{SPECTER} configurations. We find that integrating over the passbands results in a $<6\%$ decrease in SNR$_{\mu}$ for the 16-band set-up and $<2\%$ for the 34-band multichroic case. Given these results, we choose to compute and quote all forecast results in this paper with the sky model evaluated at the central frequencies of each band.

\begin{table*}
\begin{threeparttable}[t]
\cprotect\caption{\label{tab:multichroic} Sensitivities for the \emph{SPECTER} 34-band multichroic set-up\tnote{\textdagger}. The table lists band edges [GHz], central frequencies [GHz], instantanous per-detector noise [$\mu{\rm K}_{\rm CMB}\sqrt{s}$], number of detectors,  number of detectors per feedhorn, integrated noise [Jy/sr], and map-domain noise [$\mu {\rm K}_{\rm CMB}-$arcmin] assuming $t_{\rm obs}=1$ year and a full-sky observation.}
\begin{tabular}{|l|c|c|c|c|c|c|}
%\colrule
\hline
Band & $\nu_{\rm }$& NET& $N^\mathrm{det}_{\nu}$ & Detectors & $C_{\nu}$&$C_{\nu}^{\rm map}$\\
$\left[ {\rm GHz} \right]$ & $\left[ {\rm GHz} \right]$ & $\left[ \mu{\rm K}_{\rm CMB}\sqrt{\rm s} \right]$ & &per feedhorn& $\left[ 10^{-2} \,\, {\rm Jy/sr}\right]$&$\left[\mu{\rm K}_{\rm CMB}-\mathrm{arcmin}\right]$\\
%\colrule
\hline
1-2&1.5&518.5&8&4&0.2256&397.7\\
2-3&2.5&490.6&16&4&0.4191&266.0\\
3-4&3.5&500.5&24&4& 0.6843&221.6\\
4-10&7&227.1&24&4&1.241&100.6\\ 
\hline
10-10.75&10.375&265.9&\multirow{4}{*}{8}&\multirow{4}{*}{2}&5.520&204.0\\
10.75-11.5&11.125&285.3&&&6.805&218.8\\
11.5-12.25&11.875&268.8&&&7.303&206.2\\
12.25-13&12.625&290.5&&&8.915&222.8\\
\hline
13-20.5&16.75&74.69&\multirow{4}{*}{20}&\multirow{4}{*}{2}&  2.544&36.23\\
20.5-28&24.25&77.25&&&  5.472&37.47\\
28-35.5&31.75&77.92&&&9.359&37.80\\
35.5-43&39.25&78.07&&&14.14&37.87\\
\hline
43-56.5&49.75&56.03&\multirow{4}{*}{100}&\multirow{4}{*}{2}& 7.117&12.15\\
56.5-70&63.25&57.82&&& 11.42&12.54\\
70-83.5&76.75&57.70&&&  16.00&12.52\\
83.5-97&90.25&59.13&&&  21.43&12.83\\
\hline
97-115.75&106.375&50.82&\multirow{4}{*}{100}&\multirow{4}{*}{2}& 23.67&11.02\\
115.75-134.5&125.125&52.27&&& 30.33&11.34\\
134.5-153.25&143.875& 54.78 &&& 37.29&11.88\\
153.25-172&162.625&59.08&&&44.97&12.82\\
\hline
172-207.25&189.625 &46.03&\multirow{4}{*}{100}&\multirow{4}{*}{2}&38.48&9.985\\
207.25-242.5&224.875&53.78&&& 46.21&11.67\\
242.5-277.75&260.125&63.66&&& 51.72&13.81\\
277.75-313 &295.375&75.78&&&54.50&16.44\\
\hline
313-367.75&340.375&77.53&\multirow{4}{*}{100}&\multirow{4}{*}{2}&44.24&16.82\\
367.75-422.5&395.125&112.7&&& 44.40&24.45\\
422.5-477.25&449.875&163.7&&&41.29&35.52\\
477.25-532& 504.625 & 248.8&&& 37.86&53.98\\
\hline
532-757&644.5&329.7&100& 1&11.37&71.51\\
757-913&835&4671&100& 1& 15.85&1013\\
913-1174&1043.5&$5.836\times10^{4}$&100& 1&12.29&$1.266\times10^{4}$\\
1174-1501&1337.5&$4.259\times 10^{6}$&100& 1&13.66&$9.240\times10^{5}$\\
1501-1858&1679.5&$8.699\times 10^{8}$&100&1& 16.82&$1.887\times10^{8}$\\
1858-2000&1929&$2.040\times 10^{11}$&100& 1& 84.85&$4.426\times10^{10}$\\
%\colrule
\hline
\end{tabular}
\begin{tablenotes}
       \item [\textdagger] \footnotesize{i.e., the frequency set-up is the same as in Table~\ref{tab:fid_setup}, except that multichroic technology is used --- each of the bands between 10-500 GHz in the fiducial instrument are split into four equal-width bands.}
     \end{tablenotes}
\end{threeparttable}
\end{table*}

\begin{figure*}
    \centering
    \includegraphics[width=\textwidth]{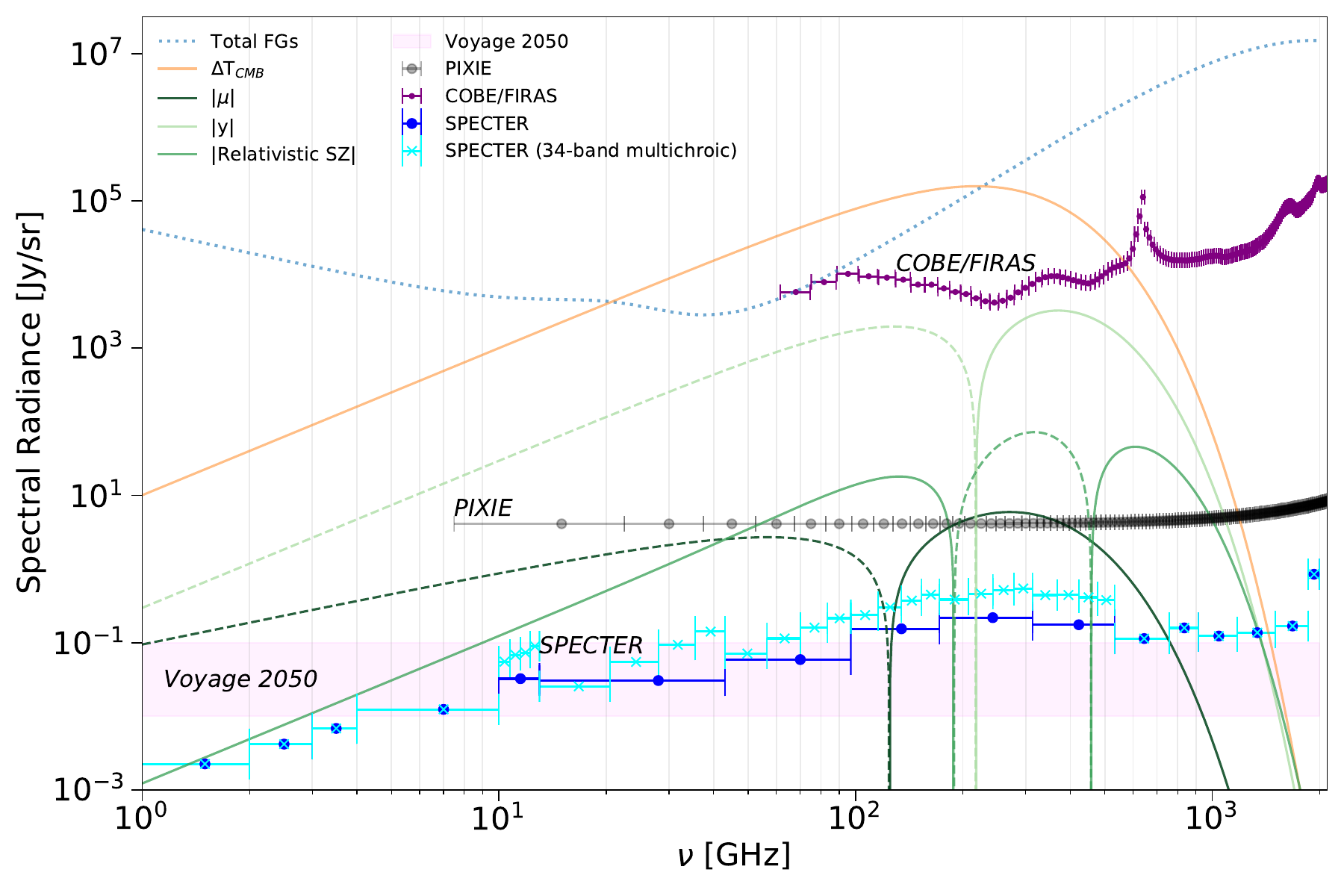}
    \cprotect\caption{\label{fig:noise_sds} \emph{SPECTER} sensitivity (blue/cyan circles with horizontal bars for optimized 16-band/34-band multichroic configuration assuming $t_{\rm obs} = 1$ year) plotted in comparison to the total foregrounds (dotted blue), CMB spectral distortions (green) including $\mu$ (darkest shade), relativistic tSZ (intermediate shade), and $y$ (lightest shade), with positive (negative) values shown in solid (dashed), and the sensitivities for several other missions: \emph{PIXIE} (dot-dashed black), \emph{Voyage 2050} (magenta), and \emph{COBE/FIRAS} (purple). We use the \emph{PIXIE} noise from A17 scaled to a 12-month duration. The \emph{SPECTER} and  \emph{PIXIE} sensitivities assume a full-sky observation here. For \emph{Voyage 2050}, we use the nominal sensitivity from Ref.~\cite{Voyage2050}.}
\end{figure*}

\section{Calibration}\label{sec:calibration}

In addition to frequency bands and detector counts, we need to determine the necessary absolute gain calibration precision for \emph{SPECTER}. To do this, we again use the Fisher formalism to estimate the impact of systematic errors on the measurement of $\mu$. 

The observed spectrum, $I_{\nu}^{\rm obs}$, will differ from the true monopole spectrum, $I_{\nu}^{\rm true}$, by some amount equal to our uncertainty in calibration. We can quantify the impact of this uncertainty via a Fisher bias calculation following Refs.~\cite{McCarthy2022,Amara2008,Huterer2005}: 

\begin{equation}\label{eq:bias}
\begin{split}
     B(p^{i})=F_{ij}^{-1}\sum_{\nu,\nu'}\frac{\partial I_{\nu}^{\rm fid}}{\partial p^{j}}C_{\nu\nu'}^{-1}\Delta I_{\nu}^{\rm sys}\, \mathrm{, and}\\
     \Delta I_{\nu}^{\rm sys}=I_{\nu}^{\rm obs}-I_{\nu}^{\rm true}
\end{split}
\end{equation}
where $\Delta I_{\nu}^{\rm sys}$ represents our systematic or calibration uncertainty, which can be either positive or negative (i.e., the spectrum can be systematically lower or higher across frequencies) and $I_{\nu}^{\rm fid}$ is the total fiducial model spectrum, which includes all the distortion and foreground signals described earlier. 

To correctly establish the calibration requirements, we need to compute the bias on $\mu$ for the different possible scenarios --- assuming the observed spectrum can be shifted up or down by some amount across either individual bands or groups of them. We first explore the scenario in which our calibration uncertainty ($\Delta I_{\nu}^{\rm sys}$ in Eq. \ref{eq:bias}) is $1/3$ of our statistical error. For each band then $\Delta I_{\nu}^{\rm sys}=\pm 1/3\times C_{\nu}$ using $f_{\rm sky}=0.7$ and for the 16-band set-up and $t_{\rm obs}=1$ year, we can compute what the bias on $\mu$ is for $2^{16}$ different cases. We find that the largest bias occurs when most neighboring bands have systematic uncertainties in opposite directions. However, the maximum bias on $\mu$ in this exploration is only $\sim 0.8\sigma$ (e.g., for dichroic calibrators --- meaning 8 calibrators in total or 1 per each neighboring pair of bands --- the bias is $\sim 0.3\sigma$). It is therefore advantageous to have fewer calibrators so that groups of bands are biased in the same way.

We repeat these computations now assuming three realistic cases in which independent calibrators are used for: 
\begin{enumerate}
    \item [(1)] each dichroic band set --- i.e., the systematic shift is the same for pairs of neighboring bands.
    \item [(2)] $\nu <40$ GHz, $\nu=40 - 400$ GHz, and $\nu \gtrsim 400$ GHz (i.e., bands 1-6, 7-10, 11-16)
    \item [(3)] $\nu <40$ GHz and $\nu > 40$ GHz (i.e., bands 1-6, 7-16)
\end{enumerate}

We assume a calibration uncertainty $\Delta I_\nu^{\rm sys}$ of either $\pm10^{-2}\,\, \mu {\rm K}_{\rm RJ}$ or $\pm10^{-3} \,\, \mu {\rm K}_{\rm RJ}$ in these scenarios (note that here we use thermodynamic temperature units in the Rayleigh-Jeans limit ($\mu {\rm K}_{\rm RJ})$; unit conversions are listed in Appendix \ref{app:units}). For the three cases, we find the bias to be $\sim65\sigma/6.5\sigma$, $\sim10\sigma/1\sigma$, and $\sim3.2\sigma/0.3\sigma$ assuming $\pm10^{-2}\,\,\mu {\rm K}_{\rm RJ}/\pm10^{-3} \,\, \mu {\rm K}_{\rm RJ}$ uncertainty, respectively. From this, we conclude that a calibration requirement of $10^{-3}\,\,\mu {\rm K}_{\rm RJ}$ with 2-3 independent calibrators is most likely needed to stay within a $<1\sigma$ bias on the observed $\mu$. 

\begin{figure*}
    \centering
    \includegraphics[width=\textwidth]{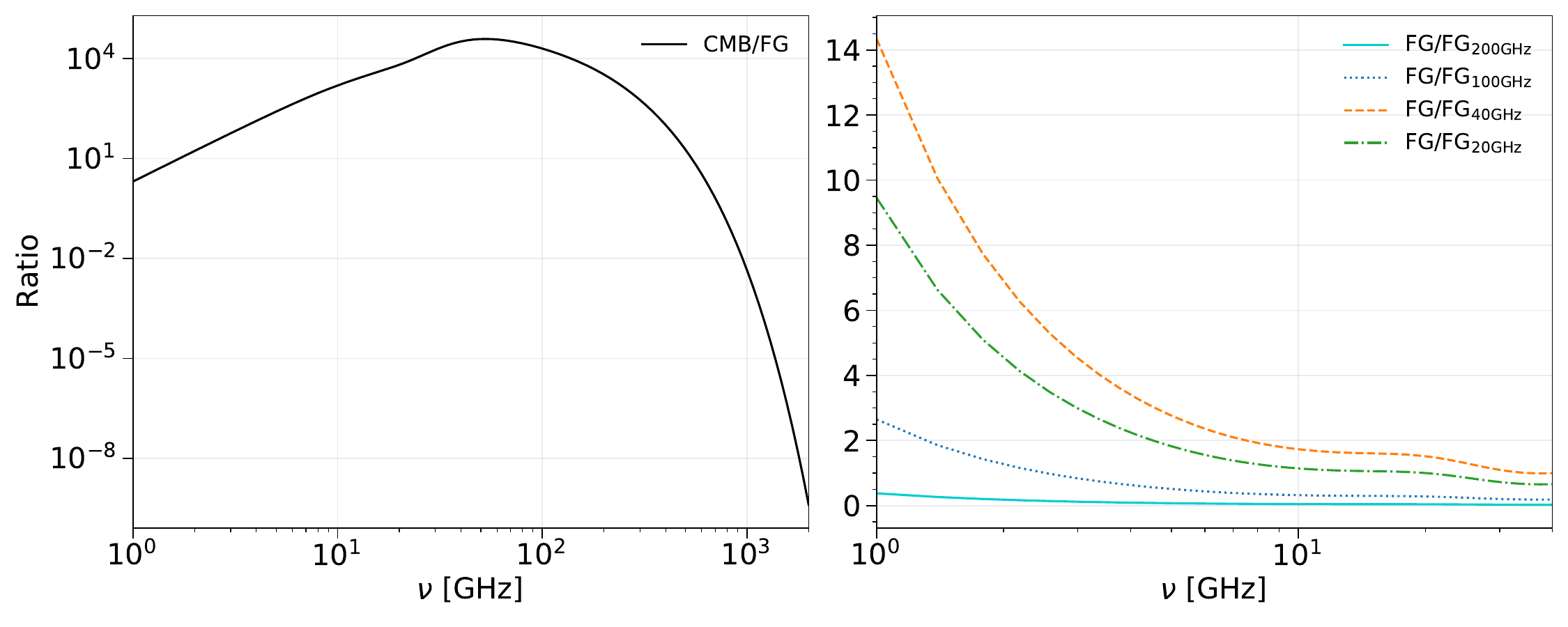}
    \caption{\emph{Left}: Ratio between the total CMB signal (including the distortions) and the total foreground signal. The CMB clearly dominates at $\sim10-800$ GHz, while foregrounds take over at $>800$ GHz. The CMB signal is still greater than the foregrounds at the lowest frequencies, but to a lesser extent than at the intermediate frequencies. \emph{Right:} The ratio of the total foregrounds between 1-40 GHz to the total foregrounds at the reference frequencies labeled in the legend. The largest difference is a factor of $\sim 10$, which suggests that calibration requirements could potentially be relaxed at the lowest frequencies as compared to the CMB-dominated frequencies, by roughly this amount.}
    \label{fig:FG_calib}
\end{figure*}
 
In the 2-calibrator set-up, we also check what happens if we use $\Delta I_\nu^{\rm sys} = 10^{-3}$ or $10^{-2}\,\,\mu {\rm K}_{\rm RJ}$ for the low and high frequencies, respectively, and find that the bias can reach $\sim 1.2\sigma$, which means that the calibration precision at the highest frequencies could be decreased. If the 2-calibrator set-up has frequency ranges 1-200 GHz and 200-2000 GHz, the higher frequencies can have a calibration precision of $10^{-2}\,\,\mu {\rm K}_{\rm RJ}$ and this would only lead to a $\sim 0.4\sigma$ bias.

To understand at which frequencies the calibration matters the most, we test three more cases for the 3-calibrator set-up (2) ($\nu<40, 40 - 400, \gtrsim400$ GHz) in which the calibration requirements are relaxed for some frequency ranges:
\begin{enumerate}
    \item [(I)] $10^{-2}/10^{-3}/10^{-2}\,\,\mu {\rm K}_{\rm RJ}$
    \item [(II)] $10^{-3}/10^{-3}/10^{-2}\,\,\mu {\rm K}_{\rm RJ}$  
    \item [(III)] $10^{-2}/10^{-3}/10^{-3}\,\,\mu {\rm K}_{\rm RJ}$
\end{enumerate}
for which we find $B(\mu)\sim 7\sigma/5\sigma/3\sigma$, respectively. This suggests that calibration precision could be possibly relaxed at the lower frequencies, where it is most challenging to achieve. We also check the use of a separate calibrator for the HEMT amplifier frequencies at $\nu < 10$ GHz (i.e., calibrators for $\nu= 1-10, 10-40, 40-400, 400-2000$ GHz or $\nu= 1-10, 10-400, 400-2000$ GHz ranges) and find that we can lower the calibration precision to $10^{-2} \,\, \mu {\rm K}_{\rm RJ}$ for the HEMT frequencies while staying within a $\sim 1.4\sigma$ bias. 

We additionally compute the bias for the 34-band multichroic configuration with the 2-calibrator set-up and find $B(\mu)\sim2.2\sigma/0.2\sigma$ for $\Delta I_\nu^{\rm sys} = 10^{-2}/10^{-3} \,\,\mu {\rm K}_{\rm RJ}$ calibration uncertainty. In the case of using a separate calibrator for the HEMT amplifiers --- a 3-calibrator set-up with $\nu= 1-10, 10-40, 40-2000$ GHz ranges --- we can lower the calibration precision for the lowest frequencies to $10^{-2} \,\,\mu {\rm K}_{\rm RJ}$, while keeping $B(\mu)$ within $0.5\sigma$. If we split the 3-calibrator set-up differently, $\nu= 1-10, 10 - 400, 400 - 2000$ GHz, we obtain $B(\mu)\sim3.7\sigma$ assuming $10^{-2} \,\,\mu {\rm K}_{\rm RJ}$ for $\nu= 1-10$ GHz. This example shows that the bias from the systematic shifts depends not only on the number of calibrators, but also on the exact frequency ranges that each of them span. This makes sense since the bias is expected to be higher in cases where a systematic shift is more degenerate with the $\mu$-distortion given the sky model used for the forecasts.

To summarize, we conclude from our bias estimates that having 2-3 calibrators at an accuracy of $\Delta I_\nu^{\rm sys} \sim 10^{-3} \,\,\mu {\rm K}_{\rm RJ}$ is important in order for the bias on $\mu$ to remain within $1\sigma$. We can decrease the calibration requirements to $\sim 10^{-2} \,\,\mu {\rm K}_{\rm RJ}$ at the HEMT amplifier frequencies, which is promising since calibration is the most difficult at the lowest-frequency bands. 

Our findings also make intuitive sense: it is not surprising that we need $\sim$nK level absolute calibration to measure the $\mu$-distortion, which has a fiducial amplitude of $\mu=2\times10^{-8}$. However, at foreground-dominated frequencies (low and high), the calibration requirements may not be as stringent, which explains why $\sim 10^{-2}\,\,\mu {\rm K}_{\rm RJ}$ precision may be sufficient at $\nu<10$ GHz. To see this, Fig.~\ref{fig:FG_calib} shows the ratio between the CMB and the total foreground signal, which thus highlights the CMB- versus foreground-dominated frequencies. We also plot the ratio of the total foregrounds between $1-40$ GHz to the total foreground at some reference frequencies. At most, it looks like the requirements could be relaxed by a factor of 10 at the lowest frequencies, similar to what we find with the bias calculations above.

In this section we have explored the maximum biases for several frequency splittings and systematic shifts to estimate the necessary calibration accuracy. We have found that a higher number of calibrators can lead to larger biases. If all bands are only allowed to coherently shift up or down by $10^{-3} (10^{-2}) \,\,\mu {\rm K}_{\rm RJ}$ (i.e., a single calibrator across all bands), then $B(\mu)\sim 0.1(1.4)\sigma$ for the 16-band and $\sim0.2(1.8)\sigma$ for the 34-band cases. Therefore, if we include some inter-band calibration in the future implementations of this instrument design, it may be possible to reduce the maximum expected bias, approaching the single calibrator limit. Finally, the calibration requirements also depend on the exact sensitivity level so we expect them to be more stringent for increased mission durations. This, again, is not unique to \emph{SPECTER} but needs to be taken into account for all future spectral distortion missions.

\section{Sky Model Robustness}\label{sec:sky_modeling_choices}
Our determination of the \emph{SPECTER} configuration relies on the fiducial sky model from A17. However, it is inevitable that the observed sky spectrum will differ to some extent from the fiducial sky model due to limitations in our current ability to model the sky. It is therefore crucial to check how the performance of our instrument configuration is affected if the sky differs from the fiducial foreground models we adopted. In other words, we want to determine how much SNR$_{\mu}$ changes if we input a different sky model in our Fisher forecast computations. To do this, we compute SNR$_{\mu}$ while varying the foreground parameters around their fiducial values within some range. To make this exploration more realistic, we vary subsets or all free parameters on a grid, thus allowing multiple parameters to vary simultaneously. The combinations where all the parameters are maximally different from their fiducial values are not necessarily the ones that would lead to the most drastic effect on the $\mu$ detection. We therefore want to ensure that we have tested other combinations of the parameter variations. 

We find that changes in the foreground spectral shape parameters (i.e., $T_{\rm d}$, $\beta_{\rm d}$, $T_{\rm CIB}$, $\beta_{\rm CIB}$, $\alpha_{\rm S}$, $\omega_{\rm S}$) have the most impact, similar to the findings in A17 for \emph{PIXIE}. This is to be expected since the Fisher forecast method, which relies on derivatives of the signal with respect to the parameters, is primarily driven by the spectral shapes of the sky model components. On the other hand, increasing or decreasing the amplitude parameters within a factor of 5 around their fiducial values has negligible effect on the forecast. Similarly, changing the values of the non-$\mu$ CMB parameters ($\Delta_T$, $y$, or $k_B T_{eSZ}$) within $\pm 20\%$ around their fiducial values does not appreciably affect the SNR$_{\mu}$ forecast (the maximum change is sub-percent).

\begin{figure}
    \centering
    \includegraphics[width=0.75\textwidth]{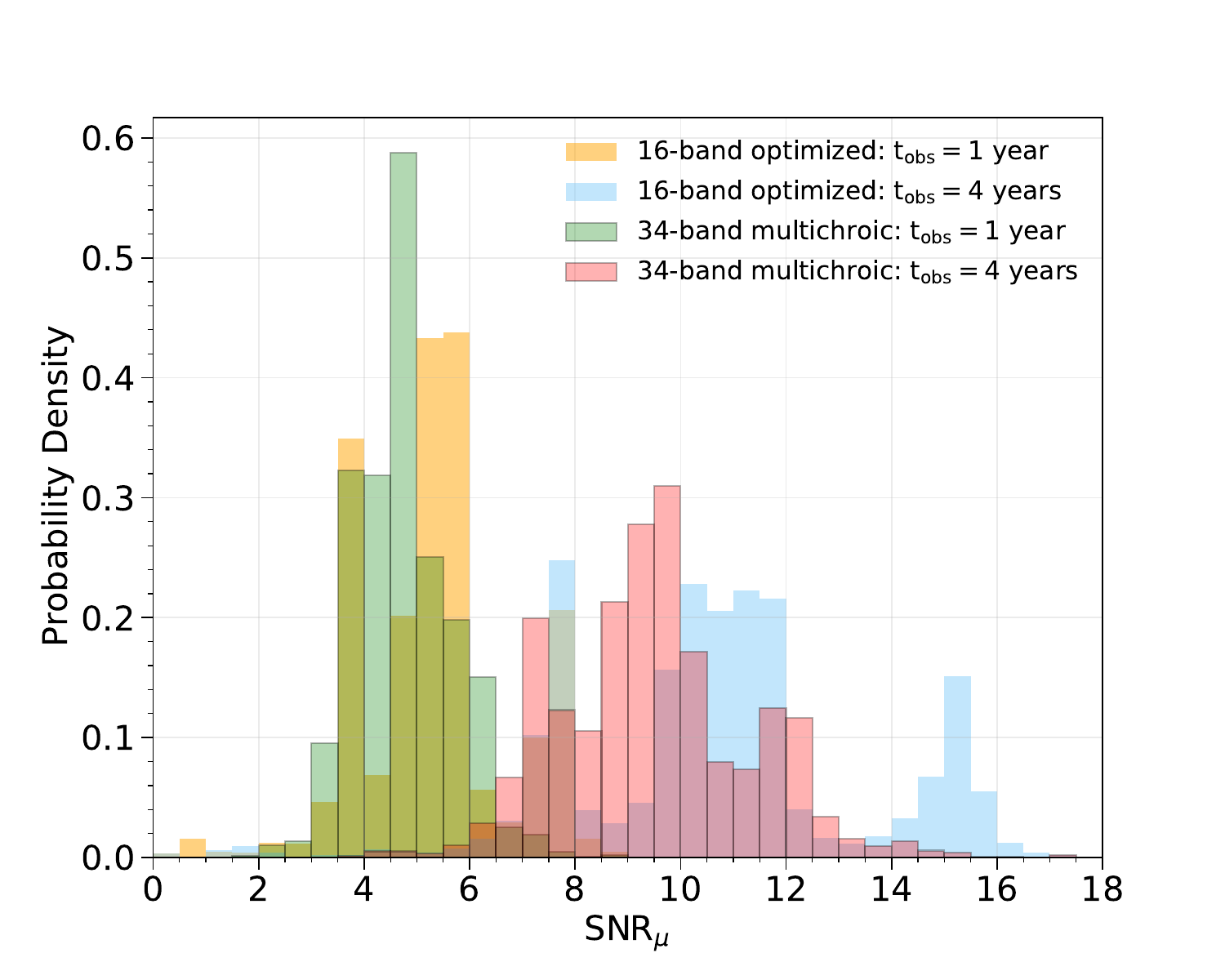}
    \caption{ 
    The effect of changing foreground spectral parameters in the Fisher forecast on SNR$_{\mu}$. The histogram demonstrates how SNR$_{\mu}$ changes if $T_{\rm d}$, $\beta_{\rm d}$, $T_{\rm CIB}$, $\beta_{\rm CIB}$, $\alpha_{\rm S}$, and $\omega_{\rm S}$ are allowed to vary within $\pm 20\%$ of their fiducial values. The results are shown for $5^{6}$ different combinations of the parameter variations --- i.e., the SNR$_{\mu}$ values are computed on a $5^{6}$-point grid, where each of the six spectral shape parameters are varied between 0.8-1.2 times their fiducial values.  We show results for the 16-band optimized configuration (orange), 34-band multichroic (green with black edges), and both of these configurations assuming $t_{\rm obs}=4$ years (blue, red with black edges) rather than 1 year. The tail at the lowest end of the SNR distribution shrinks with the additional bands included in the 34-band multichroic configuration and longer observation time. The extended 34-band multichroic mission is the most stable to foreground assumptions, with only a small set ($< 1\%$) of models yielding a $\mu$ detection at $<5\sigma$.}
    \label{fig:FG_bias_SNRs}
\end{figure}

When we allow the foreground spectral shape parameters to vary on a grid between $0.8-1.2\times$ their fiducial values, we find that in the worst cases, SNR$_{\mu}$ can decrease from the fiducial-model result of $\sim 5$ to $<1$, leading to no detection of $\mu$ for our instrument configuration, but in general, these are very uncommon (for example, in $<1\%$ of the cases for the grid calculations in Fig. \ref{fig:FG_bias_SNRs}). This is not surprising since our instrument configuration was optimized specifically in the context of our assumed sky model. However, we also find that changing the foreground model can actually result in a \emph{higher} SNR$_{\mu}$. In many cases, we find that if the observed spectrum is different from our fiducial assumption, \emph{SPECTER} could observe the $\mu$-distortion with higher significance.

As an example, in a grid calculation where we vary all spectral foreground parameters between $0.8-1.2\times$ their fiducial values with $5/10/20$ allowed values for each parameter, we find that $\sim 40-50\%$ of the scenarios lead to a decrease in SNR$_{\mu}$ (with the others yielding an increase in SNR$_{\mu}$), and only for $\sim 23-25\%$ of the cases is the decrease in detection significance greater than $1\sigma$. Thus, it is similarly likely to detect $\mu$ at a higher significance than in the fiducial forecast if the true foreground signal differs from the assumed sky model.

An example distribution of SNR$_{\mu}$ values for varied-foreground-parameter scenarios is shown in Fig.~\ref{fig:FG_bias_SNRs} (for $5^{6}$ parameter combinations). We perform the same computation for three additional instrument set-ups: the optimized 16-band configuration with $t_{\rm obs}=4$ years, the 34-band multichroic configuration (with $t_{\rm obs} = 1$ year), and the 34-band multichroic configuration with $t_{\rm obs}=4$ years. Recall that the sensitivity of the 34-band configuration is slightly lower compared to the 16-band case due to slightly lower sensitivity of the narrower channels in this configuration (see Sec.~\ref{sec:configuration_optimization} and Fig.~\ref{fig:noise_sds}). As a result, the distributions of SNR$_{\mu}$ values are shifted towards a slightly lower value for the 34-band set-ups. Fig.~\ref{fig:FG_bias_SNRs} shows that the tail of low SNR$_{\mu}$ values shrinks for the instrument with additional bands (the lowest SNR$_{\mu}\sim 1.8$ in these calculations). This conceptually makes sense: increased spectral resolution makes the configuration more robust against sky model variations. With $t_{\rm obs}=4$ years, the distribution additionally shifts towards higher SNR$_{\mu}$. In particular, for the 34-band multichroic configuration, in most cases SNR$_{\mu}$ is between 5-15 and the lowest value is $\sim3.6$ in these calculations. We note that in some sky model cases the Fisher results are not numerically stable, especially in the lowest SNR$_{\mu}$ cases ($<1$), which are essentially non-detections. There are also some cases in which the results are not stable at higher SNR$_{\mu}$, but in general these are much rarer. In follow-up work, it would be advantageous to re-parameterize the foreground models to avoid parameter degeneracies that may be causing these failures. For example, moment-based approaches offer a way to express MBB SEDs with free parameters that only have a linear dependence~\cite{Chluba2017moments}.

\begin{figure*}
\centering
\includegraphics[width=\textwidth]{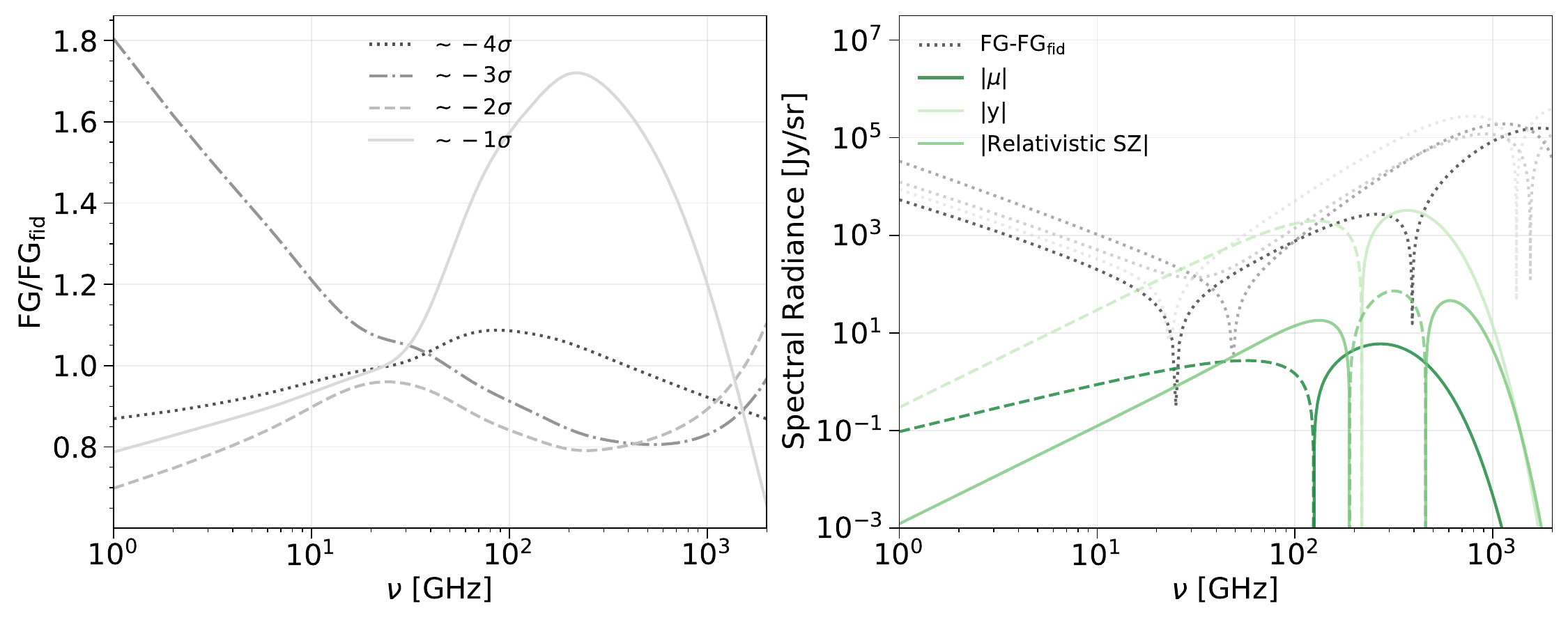}
\includegraphics[width=\textwidth]{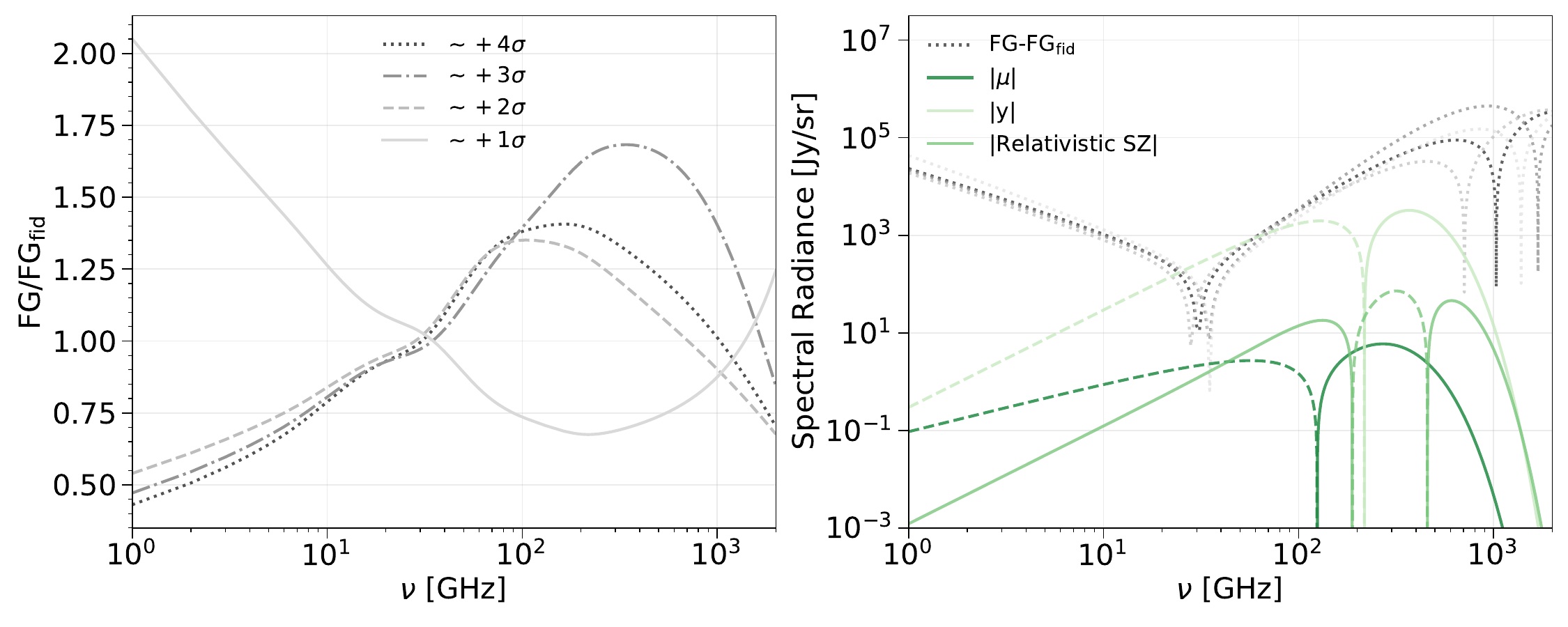}
\caption{Example foreground model variations and the corresponding impact on SNR$_{\mu}$. Upper (lower) panels show example cases that lead to a decrease (increase) in SNR$_{\mu}$, as labeled in the plot legends (e.g., ``$-4\sigma$'' means that this foreground model yields a $4\sigma$ decrease in the $\mu$ detection significance).  Plots on the left show the ratio between the total foregrounds and the fiducial foregrounds for example models chosen from a set in which $T_{\rm d}$, $\beta_{\rm d}$, $T_{\rm CIB}$, $\beta_{\rm CIB}$, $\alpha_{\rm S}$, $\omega_{\rm S}$ are allowed to vary within $\pm 20\%$ of their fiducial value. The plots on the right show the difference between the same example foregrounds shown on the left and the fiducial model. Note that the differences are several orders of magnitude larger than the expected amplitude of the $\mu$-distortion.}
\label{fig:FG_bias}
\end{figure*}

Fig.~\ref{fig:FG_bias} shows examples of the foreground scenarios that lead to the changes in SNR$_{\mu}$ discussed above. In the left panels, we show the ratio between the fiducial foregrounds and the foregrounds with spectral parameters changed by some amount. The right panels show the difference in Jy/sr along with the distortion signals for reference, showing that the foreground differences are larger than the distortion signals. We plot cases in which the foreground differences lead to a decrease or an increase in SNR$_{\mu}$. Naively, one would expect that the forecast is most negatively impacted if the foregrounds are higher than in the fiducial model at low and/or high frequencies. However, Fig.~\ref{fig:FG_bias} shows that SNR$_{\mu}$ can also be impacted in cases where the foregrounds are mostly lower or similar to those in the fiducial model. This suggests that evaluating many possible combinations of changes in the spectral parameters is the most conservative approach to checking the effects of foreground variations on SNR$_{\mu}$, as we performed above.

This exploration of foreground assumptions yields three important conclusions: (1) Our approach to the instrument configuration optimization is not entirely reliant on the fiducial sky model and can be just as or more successful in cases where the observed foregrounds differ from the assumed models. Remarkably, the chances of a higher versus lower detection significance are roughly the same within the foreground scenario space that we explore, as shown in Fig.~\ref{fig:FG_bias_SNRs}. That is to say, this instrument configuration does not represent a fine-tuned singular optimal point, which fails in every other sky model case. (2) Nonetheless, SNR$_{\mu}$ can be significantly impacted if the foregrounds differ from our fiducial model, even if they are lower in amplitude but happen to have a different spectral shape.  However, we find that scenarios where this impact completely eradicates the $\mu$ detection are very rare. (3) With increased spectral resolution and integration time, we can achieve a set-up that is more robust to the sky model and thus significantly lower the risk of a non-detection of $\mu$.

\section{SPECTER Instrument Concept} \label{sec:SPECTER}

\begin{figure*}
    \centering
    \includegraphics[width=0.9\linewidth]{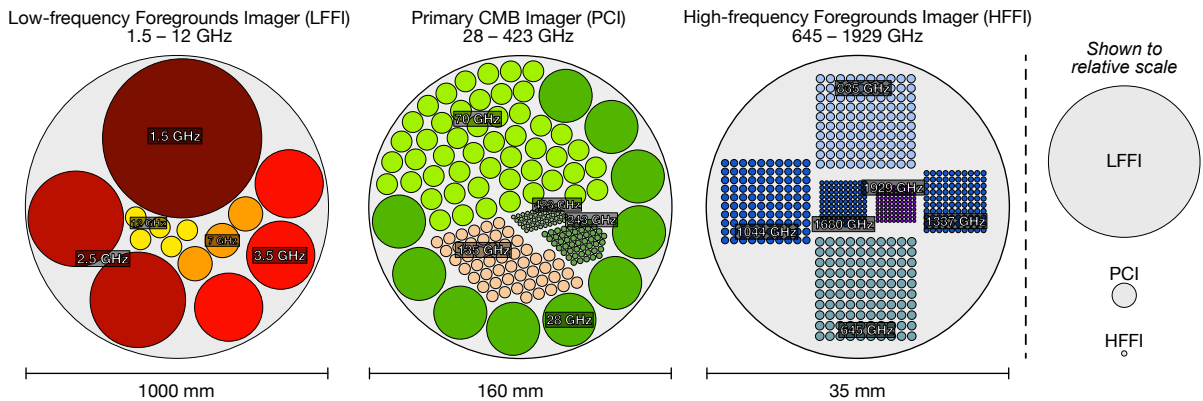}
    \caption{The 16 spectral bands optimized in Sec. \ref{sec:configuration_optimization} are distributed across four detector imaging arrays: 1) $2\times$ of the Low-Frequency Foregrounds Imager (LFFI, 1.5–12 GHz), 2) the Primary CMB Imager (PCI, 28–423 GHz), and 3) the High-Frequency Foregrounds Imager (HFFI, 645–1929 GHz). Each circle represents a feedhorn waveguide coupled to one (HFFI), two (PCI), or four (LFFI) detectors.}\label{fig:focal-plane}
\end{figure*}

In this section, we describe a concept design for \emph{SPECTER}, which implements the fiducial configuration of photometric spectral channels optimized in Sec.~\ref{sec:configuration_optimization}, as well as the calibration requirements outlined in Sec.~\ref{sec:calibration}. This instrument concept is based on proven technology that meets the requirements for our science forecasts.

\begin{figure*}
    \centering
    \includegraphics[width=\linewidth]{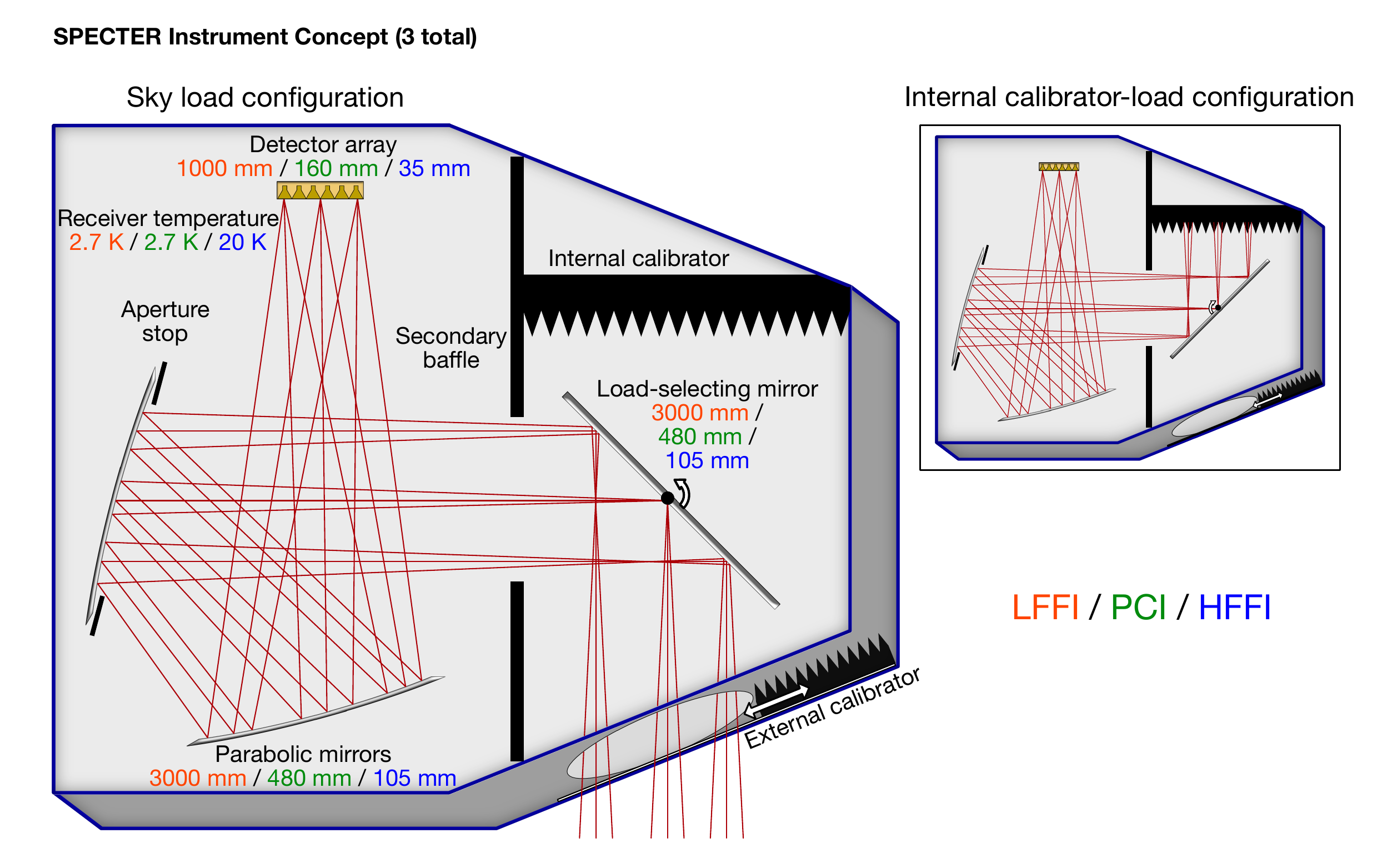}
    \caption{The complete \emph{SPECTER} instrument suite comprises four separate receiver cabins (2x LFFI, 1x PCI, 1x HFFI), which hold the detector array, coupling mirrors, aperture stop, and internal calibrator. Shown above is an example of a single receiver, along with corresponding dimensions for each instrument. In the sky-load configuration (left), the load-selecting mirror couples light from the sky to the detector array via a crossed-Dragone design comprising two parabolic mirrors. In the blackbody-load configuration (right, inset), the mirror is rotated $90\degree$ to fully illuminate the absorbing blackbody calibrator. An external calibrator, which can slide over the sky port, is added as an additional layer of redundancy. All components pictured are cryogenically cooled and maintained at 2.725~K (or 7~K for the HFFI receiver) except for the detector array, which operates at 100~mK.}
    \label{fig:instrument-design}
\end{figure*}

The design of \emph{SPECTER} is split into four total imaging focal plane arrays, as shown in Fig.~\ref{fig:focal-plane}: two Low-Frequency Foregrounds Imager (2x LFFI, band centers from 1.5–12 GHz), one Primary CMB Imager (1x PCI, band centers from 28–423 GHz), and one High-Frequency Foregrounds Imager (1x HFFI, band centers from 645–1929 GHz). This three-way split is motivated by the large variance in physical size and calibration constraints across the 1-2000~GHz bandwidth of \emph{SPECTER}. Each detector imaging array
is cryogenically cooled to 100~mK and integrated into dedicated receiver cabins that contain the telescope optics and calibration hardware (Fig.~\ref{fig:instrument-design}).

The optical design of each receiver is based on an off-axis crossed Dragone telescope comprising two parabolic mirrors, plus an additional ``load-selecting'' flat mirror. In this preliminary design, the optical system has a linear dimension approximately three times the dimension of the focal plane, making this system physically large at low frequency. We anticipate that the optimization of the optical design would prioritize the control of the illumination of the calibration load over angular resolution. Given the size of the mirror systems in this conceptual design, it is reasonable to assume the optics of the single-moded instruments would result in a resolution at the degree scale at $\sim150$ GHz, and scaling up to about 5 degrees at 1.5~GHz.

The load-selecting mirror can rotate on a flexure bearing positioned at its center axis, allowing the detector beams to alternately observe between the sky and calibration hardware. The calibration hardware includes (a) a beam-filling reference blackbody with exquisite temperature accuracy for calibrating the absolute gain of each detector channel, and (b) a frequency-tunable laser spectrometer for calibrating the passbands of the detector channels (not shown). An aperture stop at the primary mirror controls the incoming beam, while a secondary baffle between the load-selecting mirror and the primary mirror is used to further reduce the impact of stray light. 

To mitigate thermal systematics, the receiver cabin and its internal components, including the telescope mirrors, aperture stop, and calibration hardware, are cooled and stabilized to a common temperature. This approach was first demonstrated by the ARCADE 2 instrument~\cite{Singal_2011} as a strategy for significantly reducing systematics in the absolute calibration of the detectors. For the LFFI and PCI, the internal receiver temperature is 2.725~K to match the expected sky temperature. For the HFFI, the internal temperature is set to a higher 7~K to match the spectral intensity of the high frequency sky.

\subsection{Detectors}
As described in Sec.~\ref{sec:bolocalc}, the baseline detector technologies for \emph{SPECTER} are HEMT amplifier-based radiometers for the LFFI, transition-edge sensor (TES) bolometers for the PCI, and direct-absorbing bolometers for the HFFI. All three instruments use conical feedhorn arrays to couple light from the sky onto the detectors. For the LFFI and PCI, we assume two detector channels per feedhorn enabled via orthomode transducers (OMT), which split the incoming signal into orthogonal polarizations. For the HFFI, we baseline a single channel per feedhorn. All three instruments are designed to be single-moded to produce a simpler response with reduced systematics compared to an equivalent multi-moded beam.

% The LFFI radiometer architecture is based on correlated polarimeters utilizing an internal load to stabilize $1/f$ noise effects, similar to designs that have been demonstrated on previous experiments including ARCADE 2~\cite{Singal_2011}, \emph{Planck} LFI~\cite{Bersanelli_2010}, CAPMAP~\cite{Barkats_2005}, and QUIJOTE~\cite{quijote}. In this general design, radiation from the sky couples via a feedhorn waveguide to an OMT, which splits the incoming signal into two orthogonal polarization states. The signal from each polarization arm is amplified using cryogenic HEMT low-noise amplifiers and then bandpass filtered. A mixer and local oscillator may be used to down convert the signal to an intermediate frequency for improved processing. The signal is then amplified once more and read out at the output of a detector diode. To reduce $1/f$ gain effects, a switch at the input of each initial amplifier alternates between the horn-coupled OMT and a temperature-controlled internal load. 
The LFFI radiometer architecture is based on the differential radiometer design used for \emph{WMAP} \cite{Jarosik_2003}, but designed to compare a feedhorn-coupled signal against an internal reference load. In the envisioned implementation, the two polarization states emerging at the output of a given feedhorn are separated by an OMT. Each of these signals is fed into independent pseudo-correlation radiometers. The two inputs for each radiometer are the sky signal (single polarization) ($A$) and a reference thermal load tuned to match the average sky signal ($B$) with appropriate long time-scale stability. These signals enter a hybrid tee, which outputs the sum and difference of these signals. The sum and difference are then amplified by factors of $g_1$ and $g_2$, respectively, and then pass through phase switches that modulate the relative path length between the arms, corresponding to a relative factor of $\pm 1$ between the arms when switched. Demodulating the signal at each detector yields the difference between the feedhorn and reference loads with one diode measuring the signal and the other the load, but switching when the phase switch changes state. Provided the phase modulation is sufficiently fast, this also suppresses low-frequency noise, yielding a measurement at each detector diode with sensitivity given by Eq. \ref{eq:LFFI_delta_T}. 
% $\Delta T = 2 \frac{T_{\mathrm{sys}}}{\sqrt{\eta \Delta \nu}}$ (Eq. \ref{eq:LFFI_delta_T}). 

% Since each polarization is measured by two detector diodes, this leads to a sensitivity per polarization channel of: \begin{equation} \Delta T = \sqrt{2} \frac{T_{\mathrm{sys}}}{\sqrt{\eta \Delta \nu}}. \end{equation}

As an aside, we note that the sensitivity of the LFFI detectors is primarily limited by the noise temperature of the HEMT amplifiers, which operate at levels more than an order of magnitude above the quantum noise limit at these frequencies. In contrast, superconducting parametric amplifiers—currently under development~\cite{eom2012widebandlownoisesuperconductingamplifier,faramarzi202448ghzkineticinductance,howe2025compactsuperconductingkineticinductance}—have demonstrated noise performance approaching the quantum limit. Integrating these amplifiers into \emph{SPECTER} could improve the per-detector sensitivity of the LFFI by a factor of 10 or more, thereby enabling a substantial reduction in the instrument’s overall footprint.

For the PCI, a feedhorn waveguide couples to a planar OMT assembled on a detector wafer. The signal from each OMT fin feeds to a coplanar waveguide before transitioning to superconducting microstrip lines. Next, a bandpass filter made of short-circuited resonant stubs defines the passband of the detector channel. A hybrid tee then combines signals from opposing OMT fins; the difference output corresponding to the TE11 mode is directed to the TES bolometer while the sum output corresponding to higher-order modes is discarded on a termination resistor. This detector architecture has been demonstrated on several ground-based observatories including the Atacama Cosmology Telescope~\cite{Datta_2014} and the Simons Observatory~\cite{Walker_2020}.

For the HFFI, we baseline feedhorn waveguides coupled to absorbing silicon nitride membranes and thermally linked to neutron transmutation doped (NTD) germanium thermistors. This ``spider web'' bolometer design~\cite{Bock1995} has been previously demonstrated to have good on-sky performance at sub-mm and far-IR wavelengths~\cite{Holmes:08, Griffin_2010}. The passbands of the HFFI channels are set by a combination of the feedhorn waveguide and a set of metal mesh low-pass filters~\cite{ade_2006} positioned directly in front of each feedhorn array. We note that the HFFI is not sensitive to polarization in this fiducial design. However, an approach similar to the SPICA polarimeter design~\cite{spica_b-bop} could be considered in the future as an option to enable polarimetry at these high frequencies. Similarly, we note that microwave kinetic inductance detectors (MKIDs) are a rapidly maturing technology for sub-mm/far-IR observations~\cite{Baselmans_2022, chapman2022ccatprime850ghzcamera} and could also be considered as a potential alternative with several performance advantages over the baseline bolometer design.

In the final instrument design, we use 40 HEMT amplifiers instead of 8 bolometers for the 12 GHz band in the configuration (i.e., the 12 GHz band is part of the LFFI, as shown in Fig.~\ref{fig:focal-plane}), which is advantageous to reduce the size of the PCI and avoid technical difficulties with using bolometers at such a low frequency. This increases the noise by $\approx29\%$ at that band, but has negligible ($\lesssim 1\%$) effect on SNR$_{\mu}$. We therefore quote all the forecast results assuming the set-up found via optimization and listed in Tables \ref{tab:fid_setup} and \ref{tab:multichroic} for consistency. 

\subsection{Calibration hardware}
Obtaining an absolute temperature measurement of the sky requires that the signal readout from each detector channel is referenced to a power source of known absolute temperature. As discussed in Sec.~\ref{sec:calibration}, our science forecasts place stringent requirements on the uncertainty of this calibration. On \emph{SPECTER}, absolute calibration is achieved using an absorbing blackbody load that is internal to the receiver. The calibration requirements place demanding specifications on the performance of the reference load. First, the reflectance and temperature gradient across the load must be very low. For example, if the internal temperature of the receiver is to be kept isothermal to within $\Delta T \ \mathrm{\sim 1 \ mK}$, then the reflectance of the blackbody should be less than $10^{-6}$ in order to keep stray reflections at the $\mathrm{\sim nK}$ level. Additionally, as discussed below, the temperature of the calibrator must be known to within $\mathrm{\sim 1 \ nK}$ precision and $\mathrm{\sim 1 \ \mu K}$ accuracy.

To meet these performance requirements, \emph{SPECTER} adopts the general calibrator design flown on ARCADE 2~\cite{fixsen_2006} and later proposed for \emph{PIXIE} \cite{Kogut_2020}. The blackbody load consists of absorbing cones made of Steelcast (a mixture of Stycast epoxy, alumina powder, and stainless steel powder) or a similar dielectric material. The cones are mounted to an aluminum plate, forming a blackbody absorber that is larger than the beam of the lowest frequency channel for that instrument. The LFFI and PCI internal calibrators are temperature-controlled to $T_\mathrm{CMB}=2.725$ K, matching the surface brightness of the sky at these frequencies. 

The HFFI calibrator must be set to higher temperatures to provide high-frequency photons while still matching the surface brightness of the optically-thin diffuse Galactic dust foreground. Since the dynamic range of the sky signal across the full 0.6-2 THz range would be too broad for a single calibrator temperature, the HFFI calibrator must be periodically tuned to different temperature setpoints. These temperatures, corresponding to about 3.0~K (645 and 835~GHz), 4.5~K (1044 and 1337~GHz), and 6.4~K (1680 and 1929~GHz), are sufficient to keep the dynamic range between the sky and calibrator signals to well within an order of magnitude, and thus mitigate potential detector nonlinearities. The periodic tuning would lead to an overall decrease in $t_{\rm obs}$ for the HFFI channels relative to LFFI and PCI since only 1/3 of the time will be spent observing at a suitable temperature for each of the frequency ranges. We assess the overall effect from this reduction in sensitivities by computing SNR$_{\mu}$ assuming that the noise in HFFI channels (6 highest frequencies) is increased by a factor of $\sqrt{3}$ and find that the effect is negligible for both the 16-band and 34-band configurations ($\lesssim 1\%$ change). Finally, we note that the photon noise in these channels will be larger when observing the calibrator than when observing the high-frequency sky. This higher loading is not accounted for in the HFFI sensitivity calculations, but we note that the expected impact on SNR$_{\mu}$ is relatively low: about 2\% and 7\% for the 34-band and 16-band configurations, respectively.

As outlined in Section {\ref{sec:calibration}}, the temperature of the absolute calibrator must be known to within approximately $\mathrm{1~nK}$ for the PCI and HFFI, and about 10 $\mathrm{nK}$ for the LFFI. This requirement is driven by the fact that small errors in the reference blackbody, when differenced from a perfect blackbody on the sky, produce a spectrum that is partially degenerate with the $y$- and $\mu$-distortion signals. We note that this same requirement applies to any instrument operating at this sensitivity level, regardless of whether it is implemented as an FTS or as a broadband radiometer design. Although modern temperature sensors can achieve $\mathrm{\sim nK}$ precision with sufficient integration time, reaching $\mathrm{\sim nK}$ accuracy remains a far more challenging task and is beyond the capabilities of current state-of-the-art thermometry. 

To reduce this accuracy requirement, \emph{SPECTER} adopts the approach proposed by PIXIE~\cite{Kogut_2020} of taking measurements at different load temperatures to enable a marginalization over any unknown offset in the true calibrator temperature. Varying the calibrator temperature in small steps samples different points along the differential blackbody spectrum, which helps break the degeneracy between a temperature offset and a true distortion signal. Using this method, we anticipate reducing the calibrator accuracy requirement to the $\mathrm{\sim \mu K}$ level.

To provide a stable gain calibration, the instrument beams must swap between the sky and the internal calibrator at a rate that is faster than the $1/f$ noise of the detectors. For \emph{SPECTER}, we expect a $1/f$ knee of about 10~mHz, corresponding to load swaps between sky and calibrator on $\mathrm{\sim} 1$-minute timescales. In the event that this $1/f$ performance is too optimistic, we may consider developing a continuously-rotating load-selecting mirror capable of swapping on $\mathrm{\sim} 1$-second timescales. In addition to the internal absorbing load, a second blackbody calibrator is designed to be mounted to flexure bearing stages that slide across the external sky port of the receiver cabin. This external calibrator adds an additional layer of redundancy towards achieving the demanding gain calibration requirements.

Finally, we consider the option of in-situ passband calibrations as a supplement to pre-flight passband characterizations. This optional calibration is enabled via a recent commercially available frequency-tunable laser source\footnote{https://www.toptica.com/products/terahertz-systems/frequency-domain/terascan}. This source is capable of probing the complete spectral response of \emph{SPECTER}, 1 to 2000 GHz, at a very high resolution and has already been demonstrated in precise in-lab passband measurements of CMB instruments~\cite{sierra2024}. The source can be enabled or disabled on-demand, allowing for regular and unobtrusive passband measurements of each spectral channel. Although not shown in the fiducial design in Fig.~\ref{fig:instrument-design}, this passband calibrator can be easily positioned to slide across the external sky port of the receiver in a similar manner to the external blackbody calibrator.

We note that the final instrument concept is split into three different types of imaging focal plane areas that assume different frequency ranges than those considered for the calibration requirements in Sec.~\ref{sec:calibration}. As a consistency check, we compute the bias on $\mu$ assuming three separate calibrators are used for the frequency ranges of the LFFI, PCI, and HFFI. Although we have 2 LFFI instruments, we expect that we will be able to cross-calibrate the two calibrators, which cover the exact same frequencies. Thus, we can assume a single calibrator at low frequencies in computing the expected biases from calibration offsets. We find $B(\mu)\approx1.1(11)\sigma$ for a systematic error $\Delta I_{\nu}^{\rm sys} = 10^{-3} (10^{-2}) \,\,\mu$K$_{\rm RJ}$. Relaxing the calibration uncertainty to 10$^{-2} \,\, \mu$K$_{\rm RJ}$ for LFFI, while keeping it at 10$^{-3} \,\, \mu$K$_{\rm RJ}$ for PCI and HFFI, maintains the bias at $1.5\sigma$. On the other hand, loosening the calibration precision to $10^{-2} \,\, \mu$K$_{\rm RJ}$ for PCI and HFFI leads to $6.7\sigma$ and $5\sigma$ biases on $\mu$, respectively. 

For the 34-band multichroic case, we find $3.5(35)\sigma$ assuming $10^{-3}(10^{-2})\,\, \mu$K$_{\rm RJ}$ uncertainty across all three calibrators, respectively. With uncertainty of $10^{-2} \,\, \mu$K$_{\rm RJ}$ for the LFFI, the bias is $3.9\sigma$. While the bias is higher for this exact splitting of the bands across three frequency ranges, we note that these results assume maximum calibration uncertainties (i.e., the bias is computed for $2^{3}$ possible scenarios and in the case of  $\pm10^{-3}\,\, \mu$K$_{\rm RJ}$ systematic shifts, in four out of eight cases, the bias is $<1\sigma$). As previously discussed in this section and Sec.~\ref{sec:calibration}, the bias stems from the degeneracies of the systematic shifts with the $\mu$-distortion. For example, if we move the 313-532 GHz band to HFFI, the bias using the 34-band multichroic set-up is $\approx1.2(12)\sigma$ assuming systematic shifts of $10^{-3}(10^{-2})\,\, \mu$K$_{\rm RJ}$.

%%%%%%%%%%%%%%%%%%%%%%%%%%%%%%%%%%%%%%%%%%%%%%%%%%%%%%%

\section{Discussion and Conclusion}\label{sec:conclusion}

One of the open questions in cosmology today is the origin of primordial density fluctuations, which provided the seeds for all structure formation in the Universe. The CMB $\mu$-distortion, sourced by the dissipation of these perturbations on small scales, offers a unique tool to constrain early-Universe models.

In this work, we have presented a new instrument concept --- \emph{SPECTER} --- that would measure the primordial $\mu$-distortion at $\sim5\sigma (10\sigma)$ assuming 1 (4) year(s) of spectral distortion observation time (i.e., mission duration of 2 (8) years). Recall that half of the mission duration time is spent observing the sky and the other half observing the absolute calibration source (see Sec.~\ref{sec:SPECTER} and Sec.~\ref{sec:configuration_optimization}).

This concept uses a set of HEMT amplifiers and bolometers distributed into three separate instruments, each with an absolute calibration system comprised of both an internal and an external calibrator and a load-selecting mirror to swap between the blackbody source and the sky. This allows for much greater flexibility in the exact frequencies and detector counts (and hence sensitivity) compared to, for example, FTS-based set-ups, since the choices for the exact frequency bands and the number of detectors at each band can be determined independently. We have used the Fisher formalism to optimize the choice of frequency bands and the number of detectors to target the $\mu$-distortion, while minimizing the detector array area of the instrument.  In these calculations, we have marginalized over the other predicted CMB distortion signals and astrophysical foregrounds (16 free parameters in total).  Tables~\ref{tab:forecast_1year} and~\ref{tab:forecast_4years} summarize the marginalized error bars on the CMB parameters for both the 16-band and 34-band configurations, for $t_{\rm obs} = 1$ year and $t_{\rm obs} = 4$ years, respectively.  Full triangle plots showing the marginalized 2D Fisher-matrix posteriors on all sky model parameters can be found in Appendix~\ref{app:triangle_plots}.

Our optimized configuration consists of 16 bands between 1-2000 GHz with a focal plane area of $\approx1.6$ m$^{2}$ (see Table~\ref{tab:fid_setup} and Fig.~\ref{fig:focal-plane} for more details). The experimental design proposed in this paper features two compact instruments, the PCI and HFFI, as well as two considerably larger LFFI instruments. The size of the LFFI is dominated by the lowest-frequency channels, requiring feedhorns up to $\mathrm{\sim}0.5$~m in diameter. As has been previously discussed in the literature (e.g., in A17),  sensitive observations in these lowest-frequency bands is crucial to the $\mu$ detection. Future work could explore the possibility of dropping the lowest bands in order to reduce the overall size of the instrument. This could potentially be achieved via further band optimizations or through improved knowledge of the low-frequency foregrounds from future surveys (e.g., TMS absolute measurements of the emission at 10-20 GHz or European Low Frequency Survey observations at $5-120$ GHz \cite{ELFS}).
\begin{table}[]
    \centering
    \begin{tabular}{|l|c|c|c|c|}
    %\colrule
    \hline
            \multicolumn{1}{|l|}{parameter $\&$}& \multicolumn{2}{c|}{16-band}& \multicolumn{2}{c|}{34-band multichroic}\\ 
        % \cmidrule(lr){3-4}\cmidrule(lr){5-6}
        \cline{2-5}
          fiducial value& SNR & $\sigma$ & SNR & $\sigma$\\
         %\colrule
         \hline
         $\Delta_{T}=1.2\times10^{-4}$ & 37157 &$3.2\times10^{-9}$&30378&$4.0\times10^{-9}$\\
         $\mu=2\times10^{-8}$&5 & $4.0\times10^{-9}$&4.5& $4.4\times10^{-9}$\\
         $y=1.77\times10^{-6}$&955 & $1.9\times10^{-9}$&807&$2.2\times10^{-9}$\\
         $k_{\rm B}T_{eSZ}=1.245$~keV&33&0.037&42&0.029\\
         %\colrule
         \hline
    \end{tabular}
    \caption{Forecasts for the four CMB parameters using the 16-band optimized and 34-band multichroic set-ups assuming $t_{\rm obs}=1$ year. We list the fiducial values, SNRs, and the Fisher error bars.}
    \label{tab:forecast_1year}
\end{table}

\begin{table}[]
    \centering
    \begin{tabular}{|l|c|c|c|c|}
    %\colrule
    \hline
        \multicolumn{1}{|l|}{parameter $\&$}& \multicolumn{2}{c|}{16-band}& \multicolumn{2}{c|}{34-band multichroic}\\ 
        \cline{2-5}
          fiducial value& SNR & $\sigma$ & SNR & $\sigma$\\
         %\colrule
         \hline
         $\Delta_{T}=1.2\times10^{-4}$ & 74313 &$1.6\times10^{-9}$&60757&$2.0\times10^{-9}$\\
         $\mu=2\times10^{-8}$&10 & $2.0\times10^{-9}$&9& $2.2\times10^{-9}$\\
         $y=1.77\times10^{-6}$&1911 & $9.3\times10^{-10}$&1615&$1.1\times10^{-9}$\\
         $k_{\rm B}T_{eSZ}=1.245$~keV&67&0.019&85&0.015\\
         %\colrule
         \hline
    \end{tabular}
    \caption{Same as Table~\ref{tab:forecast_1year}, but for $t_{\rm obs}=4$ years.}
    \label{tab:forecast_4years}
\end{table}

We have also investigated the absolute calibration requirements for \emph{SPECTER} to avoid biases in the $\mu$-distortion measurement. The fiducial configuration calls for $2-3$ calibrators with a maximum allowed systematic uncertainty of $10^{-3} \,\, \mu$K$_{\rm RJ}$. This requires that the reference load temperatures be known to $\mathrm{\sim1 \ nK}$ precision and $\mathrm{\sim1 \ \mu K}$ accuracy. The $\mathrm{\sim \mu K}$ accuracy requirements are beyond the capabilities of current state-of-the-art cryogenic thermometers by about two orders of magnitude. However, we note that this is a fundamental challenge of the $\mu$-distortion measurement that is common to both \emph{SPECTER} and FTS-style instruments operating at this sensitivity. Therefore, we highlight this as a critical area for further research and technical development. The calibration precision requirement can be loosened at the lowest frequencies by roughly a factor of 10, due to the relative amplitude of foregrounds to CMB at these bands. Using more calibrators, and thus allowing systematic shifts to vary across more frequencies, leads to larger biases. We also identify the absolute and relative uncertainty in the passbands as an important instrumental systematic in this measurement. Assessment of passband calibration requirements is left to future work.

Since \emph{SPECTER}'s design has been optimized for the fiducial sky model used in our Fisher forecasts, we have also checked the impact of our sky model assumptions on the performance of the instrument. For example, we computed SNR$_{\mu}$ for $5^{6}$ different sky models, where the spectral foreground parameters were allowed to vary within $\pm 20\%$. We found that although the performance of the instrument can be negatively affected if the true sky spectrum differs from our fiducial model, it is similarly likely to perform \emph{better}. This shows that our configuration is not entirely fine-tuned to the fiducial sky model. Using the 34-band multichroic set-up and longer observation time makes our configuration more robust against variations of the sky model and significantly decreases the possibility of a non-detection of $\mu$. 

With our optimized 16-band and 34-band multichroic set-ups, \emph{SPECTER} will observe the $y$-distortion at sub-percent precision (e.g., $955 (1911)\sigma$ with the 16-band set-up and $807(1614)\sigma$ with the 34-band multichroic for $t_{\rm obs}= 1 (4)$ year(s)) and will also detect the relativistic tSZ monopole at $1-3\%$ precision. Using the 34-band multichroic set-up, we check whether the instrument is sensitive to the second moment of the relativistic tSZ signal (see Eq.~4 in A17)\footnote{We use fiducial parameter values from A17.}, but find that \emph{SPECTER} is not sensitive to this higher-order moment (SNR$\sim0.6-1.2$ for our fiducial sky model). It may, however, be important to properly model this correction in the future since it has a $\sim10\%$ impact on the detection of the $\mu$-distortion.

Using the 34-band multichroic configuration and $t_{\rm obs}=4$ years, we also check the feasibility of detecting other distortion signals: the cosmic recombination radiation (CRR)~\cite{cosmospec2016} and an analogue tSZ-like distortion in the CIB, sourced by the inverse-Compton scattering of CIB photons off hot, free electrons~\cite{Sabyr2022,Acharya2023}. For the CRR signal we use a template computed with \verb|CosmoSpec| \cite{cosmospec2016}, and for the CIB distortion we use a template from Ref.~\cite{Sabyr2022}.\footnote{The inverse-Compton scattering effect in the CIB was computed using the halo model formalism with a Compton-$y$ monopole of $y=1.58 \times10^{-6}$, which is slightly lower than the fiducial value used throughout this paper, but this does not meaningfully impact our results.}  Neither of these distortions are detectable: SNR$_{\rm CRR}\sim2.6$ and SNR$_{\rm tSZ-CIB}\sim0.1$.  We note, however, that the inverse-Compton CIB distortion has a noticeable effect on the detection of the $\mu$ signal (SNR$_{\mu}$ drops by $\sim27\%$). Similar to the higher-order moments in the relativistic tSZ signal, future work should be done to ensure proper modeling of these smaller signals, since \emph{SPECTER} could have sufficient sensitivity to be impacted. Additionally, the instrument configuration could be further fine-tuned to improve sensitivity to these additional distortions of interest: the CRR is not far from being within reach of \emph{SPECTER}, so further optimization work could yield a configuration capable of detecting both $\mu$ and the CRR.  More broadly, beyond these additional signals, the current instrument design would provide low-resolution ($\sim$degree-scale) full-sky maps at a wide range of frequencies, which can be used to study the SEDs of the foreground components in detail. The LFFI and PCI detectors are also polarization-sensitive and thus \emph{SPECTER} could potentially also be used to measure B-modes, for example.   

We note, also, that in our forecasts, we only use the monopole and do not use any spatial information about the foregrounds, which would further improve the component separation. Thus, our forecasts are, in some respect, conservative, since spatial information would be helpful in more effectively mitigating foregrounds (e.g., this has been shown in the re-analysis of the \emph{COBE/FIRAS} data by Refs.~\cite{Sabyr2024, Fabbian2024}).

Finally, since the $y$-distortion is predicted to be two orders of magnitude higher than $\mu$, we could ideally have a smaller experiment that would serve as prototype to \emph{SPECTER} and target the $y$-distortion first. Such a concept would cover a narrower range of frequencies both to decrease the total focal plane area and to only make use of a single type of detector technology. In Appendix~\ref{app:y_dist}, we explore potential smaller instrument concepts that would target a $y$-distortion measurement. While in this paper we do not find a simple configuration that would easily achieve this goal, future work should be done to find such a set-up.

In summary, we have presented a novel instrument design to measure the $\mu$-distortion at high significance even after marginalizing over foregrounds, using a feasible number of detectors and frequency channels.  This concept offers a new path forward to measuring CMB spectral distortions in the coming decades. There are several steps that can be undertaken to further build on this instrument concept: 

\begin{enumerate}
    \item One of the main takeaways from this work is the determination of the frequency coverage necessary to detect the $\mu$-distortion signal to high significance. The \emph{SPECTER} concept presents an initial instrument design that implements these frequency channels at the required sensitivities. While this initial design meets baseline performance goals, there remains significant room for optimization—particularly in the low-frequency instruments (2× LFFI), which in their current configuration require a substantial cryogenic volume. One promising avenue to reducing the size of the LFFI involves the development of parametric amplifiers, which may offer an order-of-magnitude sensitivity improvement over HEMT amplifiers at these frequencies and thus enable a more compact instrument footprint. 
    \item We have determined that the absolute calibration needed to avoid a large bias on $\mu$ requires thermometry accuracy at the $\mathrm{\sim \mu K}$ level. These requirements are independent of the adopted measurement strategy at this sensitivity level and therefore present a fundamental challenge to any unbiased measurement of the $\mu$-distortion signal.
    \item In this paper we used the sky model from A17. It is important to carry out further work to improve and update the foreground models using current datasets, especially at the lowest and highest frequencies.
    \item We have shown that increasing the spectral resolution and observing time improves the robustness of the instrument to the assumed sky model. Future efforts can be made to improve the stability of the performance with respect to the sky modeling.
    \item Our forecasts are conservative in the sense that we only use the frequency dependence of the all-sky monopole signal to separate the different sky components.  The statistical anisotropy of the Galactic foregrounds could be used to further improve the component separation, e.g., by performing a pixel-space analysis and using spatial information.  This work would require full-sky simulations of all components, but could significantly improve the forecasts with no change in the instrument concept.
    \item In this work, we explored possible options for a pathfinder mission to observe the $y$-distortion alone (Appendix~\ref{app:y_dist}), while using the intermediate frequencies of the \emph{SPECTER} set-up. Future work can focus on finding a configuration that would serve both as a prototype to \emph{SPECTER} and achieve this goal.
    \item We focused on the $\mu$- and $y$-distortions in this work. Further optimization of the instrument configuration to detect additional distortion signals of interest such as the CRR lines (e.g., possibly via increased frequency resolution) would expand the scope of the science targets.
    
\end{enumerate}

\section{Open-Source Code}
\label{sec:Code}
The optimization pipeline used in this paper is available at \url{https://github.com/asabyr/specter_optimization}. Throughout this work we use a detector sensitivity calculator, \url{https://github.com/csierra2/bolocalc-space} (based on the \verb|BoloCalc| package by Ref. \cite{Hill2018}), and a Fisher forecast code, \url{https://github.com/asabyr/sd_foregrounds_optimize} (modified version of the code used in A17).

\section{Acknowledgements}

We thank Giulio Fabbian for the \emph{FIRAS} noise curve. We would also like to thank Max Lee and the anonymous referee for helpful comments on the manuscript. We thank Metin San and Daniel Herman for assistance with \verb|Zodipy|. We acknowledge computing resources from Columbia University's Shared Research Computing Facility project, which is supported by NIH Research Facility Improvement Grant 1G20RR030893-01, and associated funds from the New York State Empire State Development, Division of Science Technology and Innovation (NYSTAR) Contract C090171, both awarded April 15, 2010.  AS and JCH acknowledge support from NASA grant 80NSSC23K0463 (ADAP).  JCH also acknowledges support from NSF grant AST-2108536, NASA grant 80NSSC22K0721 (ATP), DOE HEP grant DE-SC0011941, the Sloan Foundation, and the Simons Foundation. Some of the results in this paper have been derived using the healpy and HEALPix \footnote{\url{http://healpix.sourceforge.net}} package \cite{healpy1, healpy2}. We would also like to acknowledge the use of \verb|numpy| \cite{numpy}, \verb|scipy| \cite{scipy}, \verb|pandas| \cite{pandas, pandas2}, \verb|matplotlib| \cite{matplotlib} throughout this work.

\appendix
\section{Foregrounds}\label{sec:FG}
Throughout this work we use the sky model from A17. Below we summarize the foreground signals included in our forecasts. 
\subsection{Galactic Dust}
The most dominant source of foreground contamination at high frequencies ($>100$ GHz) is the thermal dust emission within our Galaxy. The exact shape and amplitude of the SED depends on the physical characteristics of the dust grains. Empirically it has been shown that a single-temperature MBB spectrum successfully describes this cumulative emission at frequencies $< 1$ THz \cite{Planck2016FG} (e.g., spectral parameters determined by \verb|Commander| and HI-CMB cross-correlation agree point-by-point to $5-10\%$ between $100-857$ GHz \cite{Planck2016FG}). For the purposes of this forecast, we similarly use a MBB spectrum to represent the Galactic dust emission: 
\begin{equation}
\begin{split}
     I_{\rm d}=A_{\rm d}x_{\rm d}^{\beta_{\rm d}}\frac{x_{\rm d}^{3}}{e^{x_{\rm d}}-1} \quad \mathrm{where}\quad x_{\rm d}=\frac{h\nu}{k_{\rm B}T_{\rm d}.}
\end{split} 
\end{equation}
with the following fiducial values: $A_{\rm d} = 1.36\times10^{6}$~Jy/sr, $\beta_{\rm d}=1.53$, $T_{\rm d}=21$~K, adopted from Ref.~\cite{Planck2016FG}.

\subsection{Cosmic Infrared Background}
Another source of foreground emission at high frequencies is the cosmic infrared background (CIB) --- the integrated dust emission of star-forming galaxies across the observable Universe. As in the case of Galactic dust, we use an MBB spectrum to model the CIB (e.g., the \emph{Planck} \cite{PlanckCIB} CIB power spectrum analysis shows no evidence for a different CIB SED and is in good agreement with the MBB fit to the CIB mean spectrum from Ref.~\cite{Gispert2000}),
\begin{equation}
    I_{\rm CIB}=A_{\rm CIB}x_{\rm CIB}^{\beta_{\rm CIB}}\frac{x_{\rm CIB}^{3}}{e^{x_{\rm CIB}}-1} \quad\mathrm{where}\quad x_{\rm CIB}=\frac{h\nu}{k_{\rm B}T_{\rm CIB}.}
\end{equation}
with $A_{\rm CIB} = 3.46\times10^{5}$~Jy/sr, $\beta_{\rm CIB}=0.86$, and $T_{\rm d}=18.8$~K, determined by fitting \emph{Planck} data \cite{PlanckCIB}.

\subsection{Synchrotron}
Synchrotron radiation produced by the accelerated electrons in our Galaxy dominates the sky at low frequencies. The spectrum is expected to follow a power-law, while \emph{Planck} data also suggest some flattening at low frequencies \cite{Planck2016FG}. Here, we use a power-law SED with logarithmic curvature to allow for a more general spectral shape. This follows the moment expansion approach from Ref.~\cite{Chluba2017moments}, which introduced a parametric Taylor expansion as a method to account for spatial and line-of-sight averaging of the sky emission:
\begin{equation}
\begin{split}
    I^{\rm sync}_{\nu}=A_{\rm S}\left(\frac{\nu}{\nu_{0}}\right)^{\alpha_{\rm S}}\left[ 1+\frac{1}{2}\omega_{\rm S}\ln^{2}\left(\frac{\nu}{\nu_{0}}\right)\right],\\
    \nu_{0}=100 \,\mathrm{GHz}.
\end{split}
\end{equation}
Our fiducial parameters have been estimated by fitting this SED to the \emph{Planck} spectrum: overall amplitude $A_{\rm S}=288.0$~Jy/sr, spectral index $\alpha_{\rm S}=-0.82$, and logarithmic curvature index $\omega_{\rm S}=0.2$. 

\subsection{Free-free emission}
Another source of foreground emission at low frequencies is the Galactic free-free or Bremsstrahlung radiation from electron-ion scattering. For this foreground component, we use the SED model from Ref.~\cite{Draine2011}, which depends on two parameters: the emission measure and electron temperature. In the frequency range used in A17 and this work, the emission measure acts as an amplitude parameter ($A_{\rm FF}$) while the electron temperature can be fixed since the spectrum does not strongly depend on it:
\begin{equation}
\begin{split}
    I^{\rm FF}_{\nu}=A_{\rm FF}\left(1+\ln\left[1+\left(\frac{\nu_{\rm ff}}{\nu}\right)^{\sqrt{3}/\pi}\right]\right)\\ \mathrm{with}\quad
    \nu_{\rm ff}=\nu_{\rm FF}(T_{\rm e}/10^{3} \,\, \rm K)^{3/2} \\ \mathrm{and}\quad T_{\rm e}=7000 \,\, \mathrm{K}, \quad \nu_{\rm FF}=255.33 \,\, \mathrm{GHz}.
\end{split}
\end{equation}
We use a fiducial value of $A_{\rm FF}=300$ Jy/sr, which was set by fitting the free-free \emph{Planck} spectrum from Ref.~\cite{Planck2016FG} to this model.

\subsection{Cumulative CO}
One of the extragalactic foreground components at intermediate frequencies is the integrated carbon monoxide (CO) line emission from star-forming galaxies \cite{Righi2008}. To model this signal we use a template generated from the spectra in Ref.~\cite{Mashian2016} with a single free parameter: the overall amplitude $A_{\rm CO}$, with a fiducial value $A_{\rm CO}=1$:
\begin{equation}
    I^{\rm CO}_{\nu}=A_{\rm CO}\Theta^{\rm CO}_{\nu}\quad\mathrm{where}\quad \Theta^{\rm CO}_{\nu} = \mathrm{CO}\,\, \mathrm{template}_{\nu}.
\end{equation}
The spectrum of this emission is not currently well modeled, but future line-intensity-mapping experiments together with galaxy surveys are expected to significantly improve our understanding \cite{Serra2016, LIMreview, LIMreview2, Chung2022}. More information about the CO emission spectrum can also be obtained from mm-wave data via cross-correlations with the CIB \cite{Maniyar2023CO,Kokron2024}.

\subsection{Spinning Dust Grains}
Another foreground component at intermediate frequencies is the microwave emission produced by spinning dust grains with a non-zero electric dipole moment \cite{Erickson1957, DraineLazarian1998} (we denote this foreground with subscript AME for Anomalous Microwave Emission). Similarly to CO emission, we use a template, calculated from the model used by \emph{Planck} \cite{Planck2016FG}, with one free amplitude parameter $A_{\rm AME}=1$, to model this foreground:
\begin{equation}
I^{\rm AME}_{\nu}=A_{\rm AME}\Theta^{\rm AME}_{\nu}\quad\mathrm{where}\quad \Theta^{\rm AME}_{\nu} = \mathrm{AME}\,\,\mathrm{template}_{\nu}.
\end{equation}

\section{Full Fisher forecast constraints}
\label{app:triangle_plots}
\begin{figure*}
    \centering
    \includegraphics[width=\linewidth]{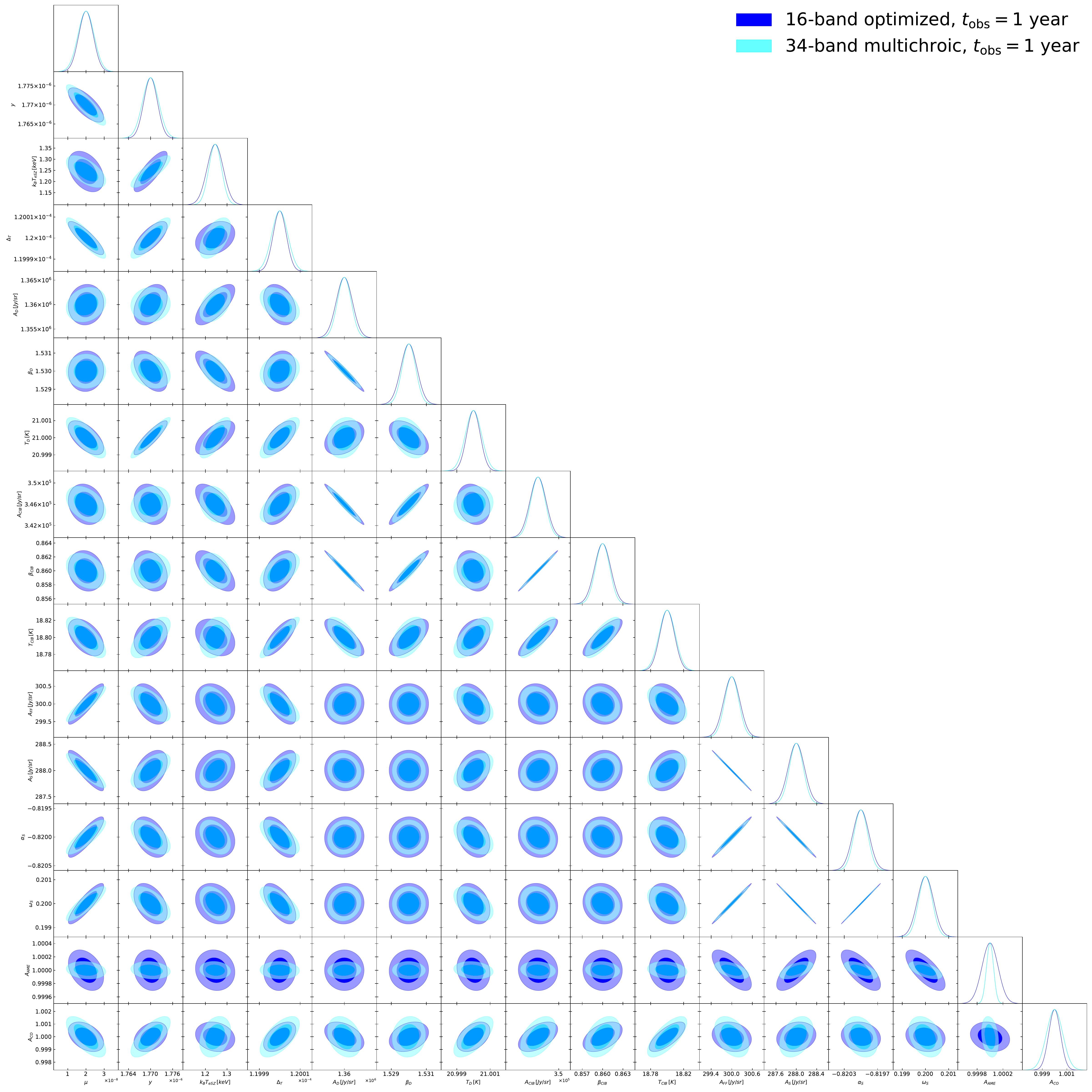}
    \caption{Parameter constraints using 16-band optimized and 34-band multichroic configurations with $t_{\rm obs}=1$ year. We show the errors on all four CMB parameters and the twelve astrophysical foreground parameters.}
    \label{fig:forecast_1year}
\end{figure*}

\begin{figure*}
    \centering
    \includegraphics[width=\linewidth]{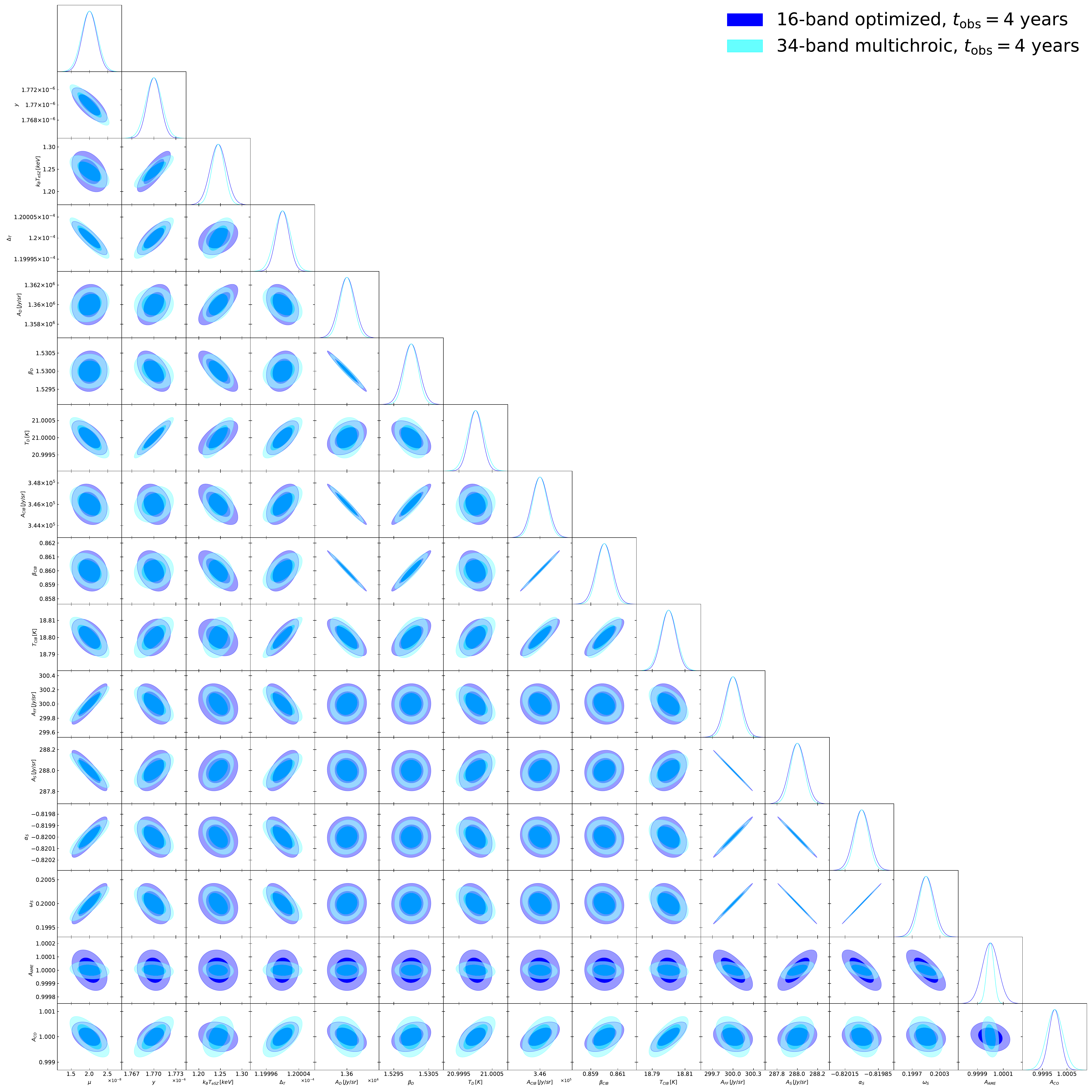}
    \caption{Same as Fig.~\ref{fig:forecast_1year}, but for $t_{\rm obs}=4$ years.}
    \label{fig:forecast_4year}
\end{figure*}

In Fig.~\ref{fig:forecast_1year} and Fig.~\ref{fig:forecast_4year}, we show the constraints on all the parameters of the sky model used in this work assuming $t_{\rm obs}=1$ year and $t_{\rm obs}=4$ years respectively, using both the 16-band optimized and the 34-band multichroic configurations.

\begin{figure*}
    \centering
    \includegraphics[width=\linewidth]{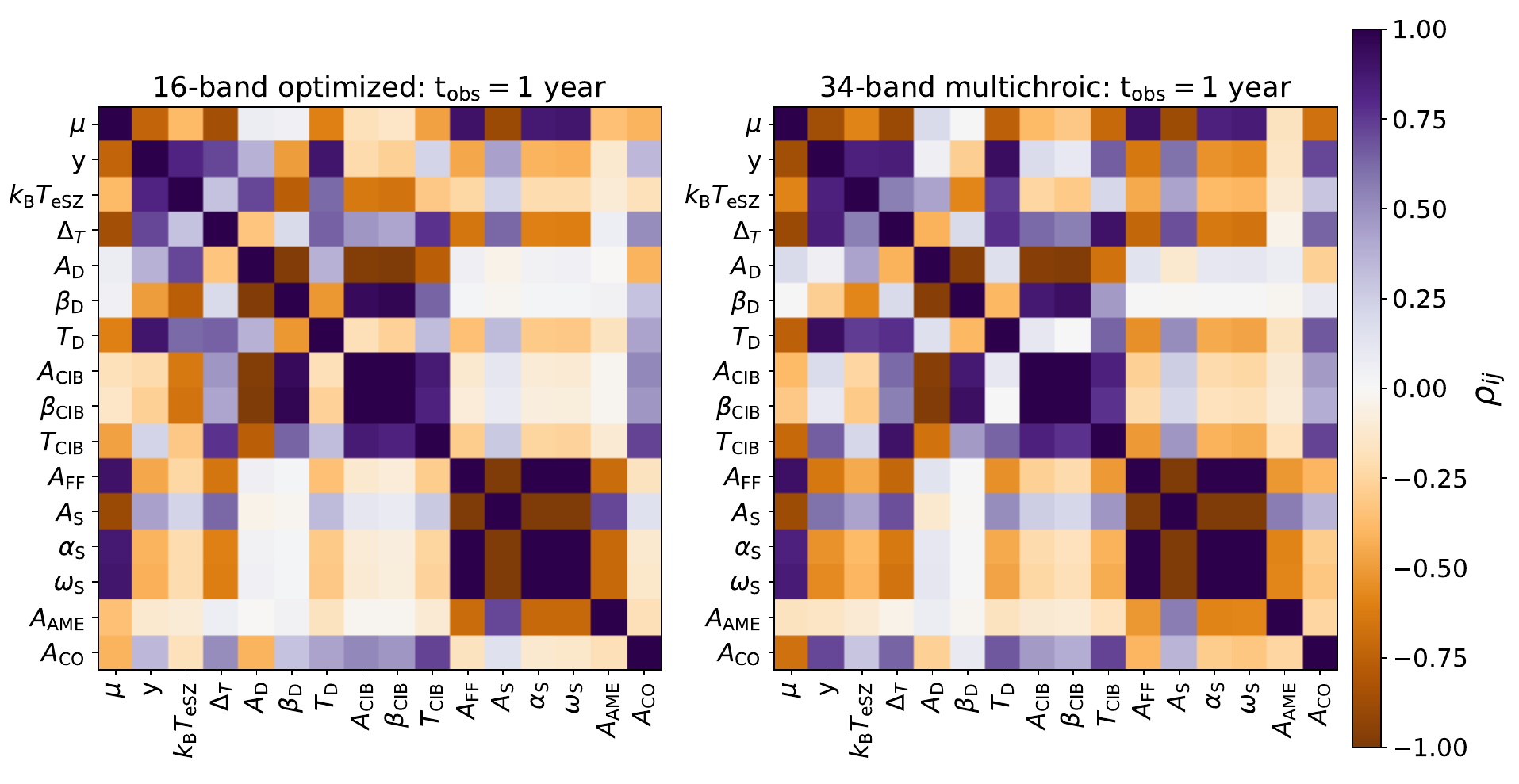}
    \caption{Parameter correlation matrices for the \textit{SPECTER} configurations. Notice the high correlation between low-frequency foreground parameters and $\mu$.}
    \label{fig:corr}
\end{figure*}

Following Ref.~\cite{Chung2024CO}, we also include the parameter correlation matrix obtained from the Fisher forecast. In Fig.~\ref{fig:corr}, we show the correlation matrices for the 16-band and 34-band configurations. We can see clearly that the low-frequency foreground parameters have the highest correlation with $\mu$, which is consistent with our findings that observing the low frequencies is essential to obtaining a measurement of the $\mu$-distortion. We can also see a high correlation among the CMB parameters, low-, and high-frequency foregrounds, as expected.

\section{Galactic Emission Lines}\label{app:emission}
\begin{figure*}
    \centering
\includegraphics[width=0.6\linewidth]{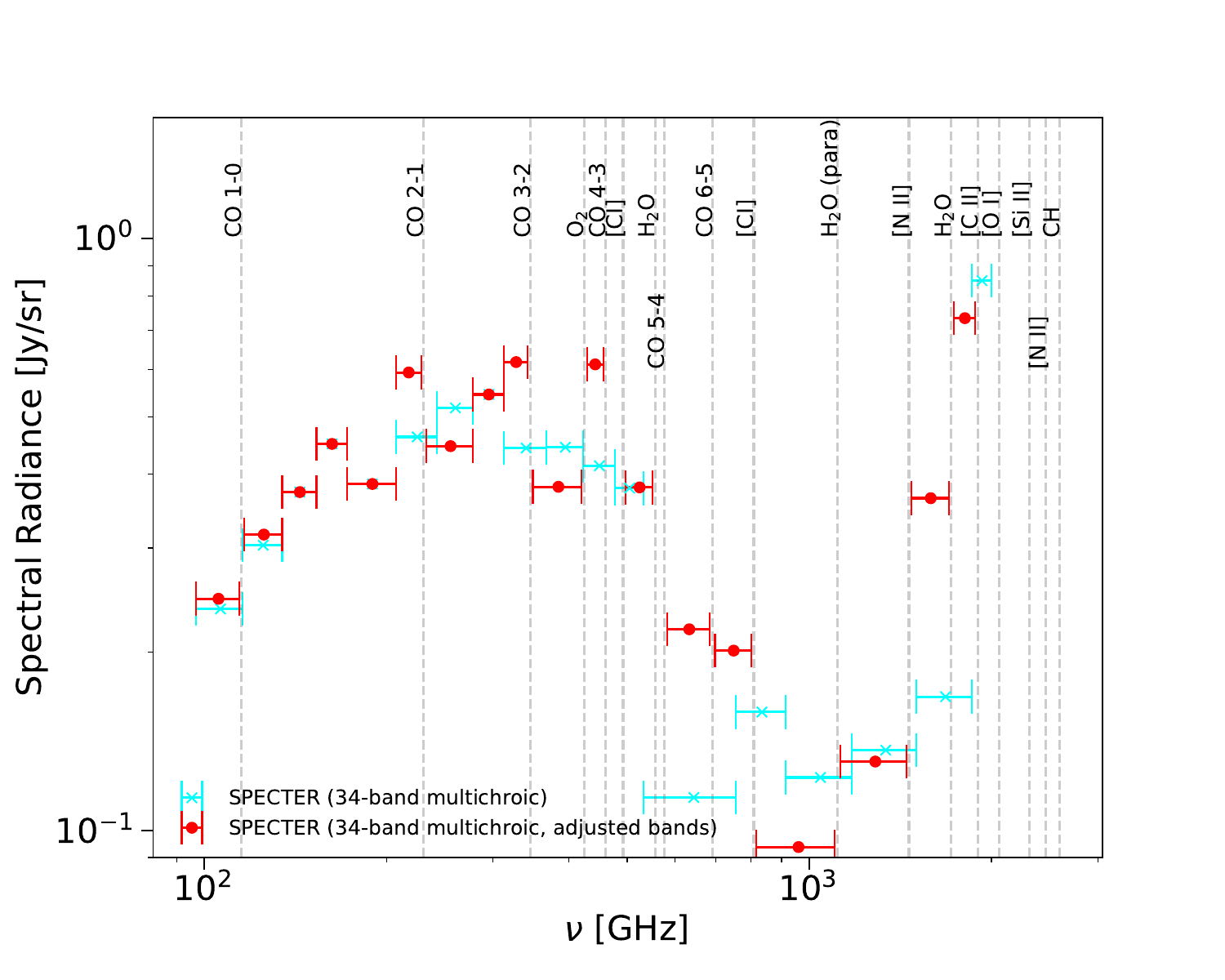}
    \caption{34-band multichoic bands (cyan cross), Galactic spectral lines based on Ref.~\cite{Fixsen1999_lines} (grey dashed), and a set of new bands with adjusted edges, which avoid the spectral lines (red circle). We show sensitivities assuming 1 year of spectral distortion integration time. Using a set of adjusted bands, which avoid Galactic spectral lines, only minimally affects SNR$_{\mu}$ ($\approx2\%$). Thus, \textit{SPECTER} can be easily built to be robust to any known Galactic spectral lines.}\label{fig:emission_lines}
\end{figure*}

We include extragalactic CO emission in all Fisher forecasts in this work. To assess the impact of Galactic spectral line emission, we consider the spectral lines provided on LAMBDA \footnote{\url{https://lambda.gsfc.nasa.gov/product/cobe/firas_linemaps.html}} \cite{Fixsen1999_lines}. The list contains spectral lines spanning the frequency range between $\approx115-2600$ GHz either detected by \textit{COBE/FIRAS} or predicted to be present in this range, including the rotational CO lines detected by \textit{Planck} \cite{PlanckCO}.

In Fig.~\ref{fig:emission_lines}, we plot all the spectral lines from the list and the subset of the bands in the 34-band multichroic set-up, which fall in the same frequency range. It can be seen clearly that several bands contain either one or several spectral features. 

One possible solution to prevent contamination from the Galactic emission lines is to construct the \textit{SPECTER} bands specifically to avoid the frequencies that are affected. For example, we can perform a test in which we assume conservatively that each spectral feature is velocity-broadened by $\pm1\%$ and adjust the band edges to avoid these frequency ranges. This level of broadening is larger than what is expected from the Doppler shift broadening from orbital velocities in the Milky Way (e.g., the Solar System's velocity of $\approx200$ km/s)
and what was assumed in the \textit{Planck} \cite{PlanckCO} analysis. 

In Fig.~\ref{fig:emission_lines}, we show a set of possible new bands with adjusted edges, which avoid the spectral lines. We extend some of the bands to fill in the additional frequency gaps formed from these adjustments. The figure shows that there is a slight loss of sensitivity for most bands, but this leads to minimal effects on SNR$_{\mu}$. Using these new bands, the $\mu$-distortion detection significance only drops by $\approx 2\%$. 

All of the affected channels are at high frequencies in the PCI or HFFI instruments, where the sizes of the feed-horn antennas are very small and, thus, we are not limited by the sensitivities (i.e., we can add additional detectors with minor impact to the overall cost of the instrument --- see Fig.~\ref{fig:focal-plane}). Therefore, an additional mitigation strategy is to increase the number of detectors in the adjusted bands. This ensures that SNR$_{\mu}$ stays roughly the same (e.g., if the number of detectors is doubled for the highest 16 frequencies, SNR$_{\mu}$ obtained using these adjusted bands with additional detectors improves by $<1\%$ compared to the fiducial set-up).

\section{Zodiacal light}\label{app:zodiacal}

In addition to Galactic thermal dust and cosmic infrared background emission at high frequencies, zodiacal light peaks in the mid-infrared. Zodiacal emission is the radiation from the reflected or re-emitted sunlight by the interplanetary dust (IPD) particles within our Solar System. It is most prominent at the mid-infrared frequencies but contributes emission from the optical to infrared wavelengths \cite{Leinert1975}. In fact, it can be observed with the naked eye in the sky at dusk and dawn near the horizon. 

In contrast to other Galactic and extra-galactic foregrounds, zodiacal emission has a temporal dependence because it varies depending on the position of the Earth within the Solar System at different times during the year. Moreover, the observed zodiacal emission depends on the exact position of the observer and, thus, differs across experiments. This offers both advantages and disadvantages to the removal of zodiacal emission from data. The unique dependence on time and position makes zodiacal emission challenging to model and requires that the modeling is done for each experiment individually. On the other hand, the extra dependencies make it easier to disentangle it from the other foregrounds.

We leave the exact modeling of the residuals expected after removing the zodiacal emission to future work since it requires knowing the precise details about the \textit{SPECTER} survey.  However, in order to approximately assess the impact of the residual zodiacal emission on the detection of the primordial $\mu$-signal, we estimate the sky-averaged zodiacal light spectrum based on previous models.

In particular, we use the \textit{Planck} 2018 \cite{planck2020zd} zodiacal light model implemented in \verb|Zodipy| \cite{San2022, zodipy}. The \textit{Planck} 2018 emission model is derived using the same analysis methods as in the \textit{Planck} 2015 data release using the zodiacal fitting procedure from Ref.~\cite{planck2013zd}. The \textit{Planck} zodiacal fits are obtained by fitting the \textit{COBE/DIRBE} IPD model, which is outlined in Ref.~\cite{Kelsall1998}, to \textit{Planck} data. The \textit{COBE/DIRBE} IPD model consists of a 3D model of the six IPD components in the Solar System: the diffuse cloud, three dust bands, the circumsolar ring, and the Earth-trailing feature, together with the associated thermal emission and scattering of sunlight from the IPD grains. The constituents of this model sum to a total of 80 free parameters. We refer the reader to Refs.~\cite{Kelsall1998, zodipy, planck2013zd} for more details.

To estimate the sky-averaged zodiacal emission, we compute the full sky zodiacal emission at center frequencies in the 34-band multichroic set-up, which are within the range of \emph{Planck} frequencies (100-857 GHz). We assume a reference position of an observer on Earth and time (January 1, 2025), and HEALPix pixelization (NSIDE=256 corresponding to angular resolution of $\approx$14 arcmin)\footnote{\url{http://healpix.sourceforge.net}}. We then compute the average emission across the pixels by first masking $30\%$ of the sky with a \textit{Planck} Galactic mask to be consistent with the assumed $f_{\rm sky}=0.7$ in the Fisher forecast calculations. We add the zodiacal emission spectrum as an additional foreground component in the sky model by using the average spectrum as a template, extrapolating to other frequencies, and parameterizing it with an amplitude parameter $A_{\rm ZE}=1.0$. While additional parameters may be necessary to describe this emission, this assumption is not overly optimistic because we do not take into account that the zodiacal emission will be subtracted in practice. This one-parameter model serves as an intermediate estimate of the impact from zodiacal light because it is conservative with respect to the overall zodiacal emission expected in the processed \textit{SPECTER} data, but likely requires additional parameters associated with the shape of the spectrum. 

\begin{figure*}
    \centering
    \includegraphics[width=\linewidth]{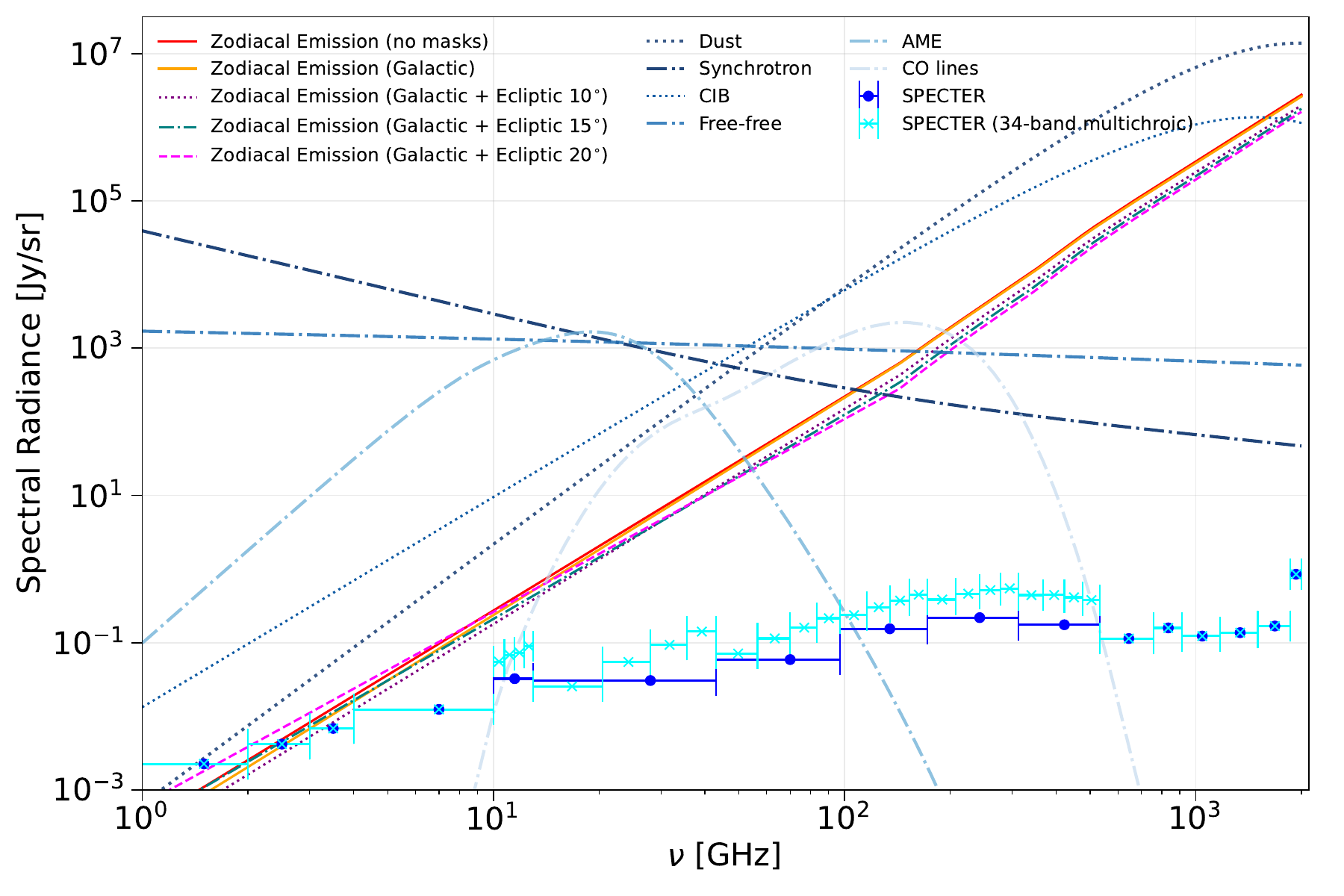}
    \cprotect\caption{Zodiacal emission in comparison to other foreground components and \textit{SPECTER} noise. We show sky-averaged zodiacal emission estimated using \verb|ZodiPy| with no sky masks applied (red solid), with a \textit{Planck} Galactic mask applied assuming $f_{\rm sky}=0.7$ (orange solid), and with additional Ecliptic masks applied, which extend 10/15/20 degrees from the Ecliptic (purple dotted/teal dash-dotted/magenta dashed).}\label{fig:zodiacal}
\end{figure*}

We find that including the zodiacal emission spectrum in the Fisher forecasts only reduces SNR$_{\mu}$ by $\approx13\%$. In Fig.~\ref{fig:zodiacal} we show the estimated zodiacal emission in comparison to other foregrounds. We can see that the zodiacal light is not dominant at any of the frequencies and peaks at the highest frequencies of \textit{SPECTER}. In the most recent \textit{Planck} maps \cite{planck2020zd}, it was shown that after modeling and subtracting zodiacal emission, the residuals were found to be neglible at 545 GHz and lower frequencies and at a low level ($10^{-4}$ Jy/sr) at 857 GHz. Thus, we can anticipate that any effects from this additional foreground can be mitigated via increased sensitivity at the high frequencies, or additional frequency channels. The size of the feedhorn antennas to which the detectors are coupled is very small at high frequencies so this does not pose risk of having to significantly increase the cost of the instrument. As an example, if we double the number of detectors at the frequency channels $>100$ GHz, the change in SNR$_{\mu}$ is $\sim9\%$.

We experiment with reducing the zodiacal emission by applying an additional mask since the zodiacal light is most prominent in the region near the Ecliptic plane. In Fig.~\ref{fig:zodiacal}, we show the average SED computed from the full sky, after applying a Galactic mask, and after including both a Galactic mask and an Ecliptic mask, which extends $10^{\circ}/15^{\circ}/20^{\circ}$ from the Ecliptic. We find that although the zodiacal emission is substantially suppressed when we mask out the region near the Ecliptic, SNR$_{\mu}$ worsens by $\approx22/26/31\%$ compared to SNR$_{\mu}$ when zodiacal light is not included. This is a more substantial effect than in the case of using only the Galactic mask previously discussed, which led to $\approx13\%$ decrease in SNR$_{\mu}$. This indicates that the increase in noise from masking a larger region of the sky is larger than the gain of suppressing the zodiacal emission.

\section{$y$-distortion experiment}\label{app:y_dist}
The amplitude of the $y$-distortion is predicted to be roughly two orders of magnitude higher than the primordial $\mu$-distortion (and about an order of magnitude below the sensitivity of \emph{COBE/FIRAS} --- see Fig.~\ref{fig:noise_sds}). With the 16-band optimized configuration and the 34-band multichroic configuration using $t_{\rm obs}=1$ year, we expect \emph{SPECTER} to measure the $y$-distortion at $955\sigma$ and $807\sigma$, respectively, while marginalizing over all other signals used in the Fisher forecast. Since measuring the $y$-distortion requires significantly less sensitivity than $\mu$, we investigate a potential design for a smaller instrument that could detect the $y$-distortion while serving as a prototype to \emph{SPECTER}.

With this in mind, we explore potential configurations for a ground-based instrument ($t_{\rm obs}=6$ months, observing a sky patch similar to SPT \cite{SPT}), a balloon-borne experiment ($t_{\rm obs}=2$ weeks, observing a similar patch of the sky as SPIDER \cite{spider}) and a SmallSat pathfinder ($t_{\rm obs}=12$ months, full sky). In particular, we investigate using a narrow range of intermediate frequencies in order to (1) limit the instrument to a single type of detector technology, (2) avoid large atmospheric and foreground contamination at higher frequencies, and (3) keep the size of the instrument small by excluding the lowest frequencies used in \emph{SPECTER}. 

\begin{figure*}
    \centering
    \includegraphics[width=\textwidth]{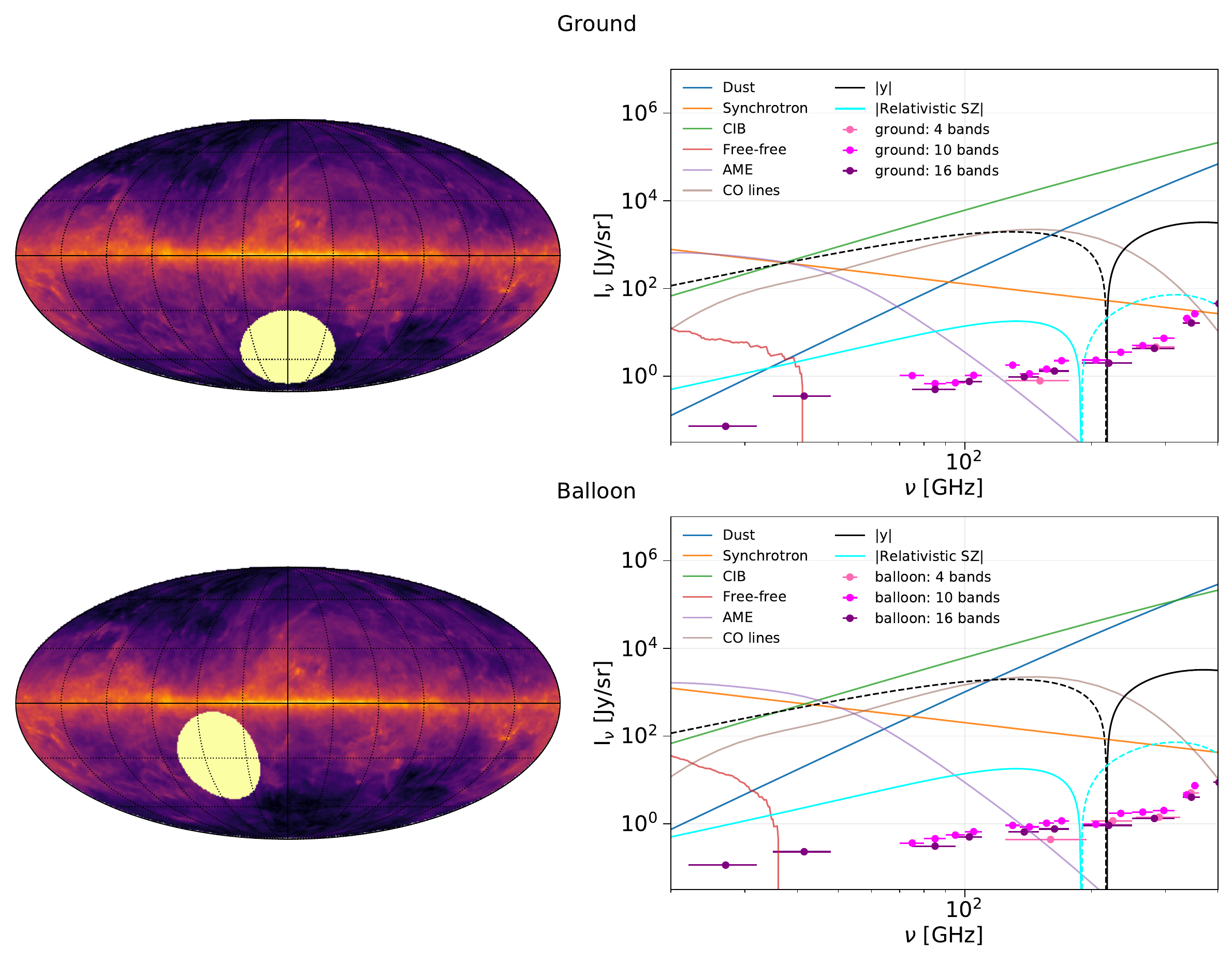}
    \caption{\emph{Left:} Sky patches used for possible $y$-distortion ground-based (top) and balloon-borne (bottom) experiments motivated by the SPT and SPIDER footprints, respectively. The map in these images shows $\ln(I_{\nu}^{d})$ computed using \emph{Planck} data, solely for the purposes of visualization. \emph{Right:} Average foreground spectra within the sky patches indicated in the plots on the left, with example instrument sensitivities and the distortion signals.}
    \label{fig:ground_balloon}
\end{figure*}

More specifically, for the ground-based and balloon-borne experiments, we consider 4-, 10-, and 16-band configurations spanning a maximum range of $\sim20-400$ GHz or narrower (see Fig.~\ref{fig:ground_balloon}). In these cases we additionally need to take into account the atmosphere, which is not included in the sky model used in the Fisher forecasts or the sensitivity estimates for \emph{SPECTER} described in the main text. To do this, we use the \texttt{am}~\cite{paine_2023_8161272} software tool to generate models of the expected atmospheric emission from both the South Pole (ground-based) and at a 40~km altitude (balloon-borne).

With these instrument configurations, the $y$-distortion is not detected when all sky signals are included in the forecast. Since in these cases one would observe a relatively ``clean'' patch of the sky (away from the Galactic plane), we look into the possibility that some of the foreground components may not be observable and could thus be excluded from the forecast. To estimate the contributions of the Galactic foregrounds in a given patch of the sky, we use \emph{Planck} foreground parameter maps and models (AME, synchrotron, and free-free maps from Ref.~\cite{Planck2016FG} and Galactic dust maps from Ref.~\cite{Planck2016CIB}). We compute the average spectra across pixels within our chosen patches of the sky for each of these Galactic components. The CIB and CO radiation have extragalactic origin and are therefore statistically isotropic across the sky.

In Fig.~\ref{fig:ground_balloon} we show the results for two example patches of the sky along with potential sensitivities for ground-based and balloon-borne experiments. The left panels show example sky regions used to estimate the foreground contributions, together with the natural log of dust emission for visual reference. We use circular areas with 25 degree radii centered within the following coordinates in right ascension (RA) and declination (DEC): -60$\degree$ < RA < 60 $\degree$ and -70 $\degree$ < DEC < -40 $\degree$ (ground-based), 20$\degree$ < RA < 80$\degree$ and -45$\degree$ < DEC < -15$\degree$ (balloon-borne). The plots on the right show the average foreground spectra within those sky patches together with the $y$-distortion, its relativistic correction, and several potential instrument sensitivities.

Based on these estimates, we cannot exclude any of the foregrounds in the forecasts since all components are above the instruments' noise levels considered here. A wider frequency coverage is most likely needed to measure the $y$-distortion at high significance. We try imposing a $7\%$ prior on $\beta_{\rm d}$ and 10$\%$ priors on the synchrotron amplitude and spectral index, but find that the detection significance is $<5\sigma$ even for the most promising instrument configurations when 3-4 or more foregrounds are included in the forecast. It is possible that there exists an optimal point where the experiment's sensitivity is high enough to observe the $y$-distortion but sufficiently low to not be sensitive to some foregrounds. We leave finding this optimal configuration to future work. 

Finally, using the same optimization approach as described for the $\mu$-distortion earlier, we also look for a possible SmallSat configuration that would only use the intermediate frequencies and observe for 12 months. However, the configurations do not provide enough sensitivity and we find their results to not be numerically stable, possibly due to the narrow frequency coverage (e.g., optimizing ranges $40-400/600/800$ GHz). We again leave further exploration of this instrument set-up to future work.

\section{Noise Units and Conversions}\label{app:units}

To obtain the conversion factor between $\mu$K and Jy/sr in Eq.~\ref{eq:noise}, we use the Planck function in Jy/sr (1 Jy $= 10^{-26}$ W m$^{-2}$ Hz$^{-1}$):

\begin{equation}
\begin{split}
    B_{\nu}(T)=\frac{2h\nu^3}{c^2}\frac{1}{e^{x}-1} \quad
    \mathrm{where}\quad x=\frac{h\nu}{k_{\rm B}T}\\
\end{split}
\end{equation}
\noindent and take its derivative with respect to temperature:

\begin{equation}
\frac{\partial B_{\nu}(T)}{\partial T}=\frac{2 h\nu^3}{c^2}\frac{e^{x}}{(e^{x}-1)^{2}}\frac{x}{T} \,.
% {\left(T_{\mathrm{CMB}}/\mu K_{CMB}\right)10^{6}}
\end{equation}
\noindent Note that this unit conversion is a function of frequency and is dependent on the temperature, by definition. 

In our assessment of calibration requirements, we use thermodynamic temperature units in the Rayleigh–Jeans (RJ) limit, denoted as $\mathrm{K}_{\rm RJ}$. This unit is independent of temperature. To convert between intensity and $\mathrm{K}_{\rm RJ}$, we use the derivative of the Planck function in the RJ limit:

\begin{equation}
\begin{split}
    B_{\nu, \rm RJ}(T) = \frac{2\nu^{2}k_{\rm B}T}{c^{2}} \quad \mathrm{and} \quad
    \frac{\partial B_{\nu, \rm RJ}}{\partial T}=\frac{2\nu^{2}k_{\rm B}}{c^{2}}
\end{split}
\end{equation}

% \nocite{*}
\bibliography{main}% Produces the bibliography via BibTeX.

\end{document}